\documentstyle[preprint,prb,aps,epsf,eqsecnum]{revtex}
\begin{document}
\tightenlines

\title{The $sl_2$ loop algebra symmetry of 
the six-vertex model at roots of unity}

\author{Tetsuo Deguchi
\footnote{e-mail deguchi@phys.ocha.ac.jp}}
\address{ Department of Physics, Ochanomizu University, 2-1-1
Ohtsuka, Bunkyo-ku,Tokyo 112-8610}
\author{Klaus Fabricius
\footnote{e-mail Klaus.Fabricius@theorie.physik.uni-wuppertal.de}}
\address{ Physics Department, University of Wuppertal, 
42097 Wuppertal, Germany}
\author{Barry~M.~McCoy
\footnote{e-mail mccoy@insti.physics.sunysb.edu}}               
\address{ Institute for Theoretical Physics, State University of New York,
 Stony Brook,  NY 11794-3840}
\date{\today}
\preprint{YITPSB-99-66}

\maketitle

\begin{abstract}

We demonstrate that the six vertex model (XXZ spin chain) with
$\Delta=(q+q^{-1})/2$ and $q^{2N}=1$
  has an invariance under the loop algebra of $sl_2$ which produces
a special set of degenerate eigenvalues. For $\Delta=0$ we compute
the multiplicity of the degeneracies using Jordan Wigner techniques

\end{abstract}
\section{Introduction}

The free energy of the six vertex model \cite{lieba}--\cite{syy} 
and the eigenvalues of the related XXZ spin
chain \cite{bethe}--\cite{yy}
 specified by
\begin{equation}
H={1\over
2}\sum_{j=1}^L(\sigma_j^x\sigma^x_{j+1}+\sigma^y_j\sigma^y_{j+1}
+\Delta\sigma^z_j\sigma^z_{j+1})
\label{hxxz}
\end{equation}
have been studied for many decades by means of Bethe's Ansatz
\cite{bethe}.
These results obtained from Bethe's Ansatz were shown by Baxter
\cite{baxa},\cite{baxb}
to
follow from functional equations which closely follow from the
star triangle equations and commuting transfer matrices.
However, it has also been shown by Baxter \cite{baxc}-\cite{baxe} 
that if we write
\begin{equation}
\Delta={1\over 2}(q+q^{-1}) 
\label{qdef}
\end{equation}
and let
\begin{equation}
q^{2N}=1
\label{root}
\end{equation}
then there are  additional properties of the model which do not follow from
the star triangle equation and commuting transfer matrices alone. Some of
these additional properties have been exploited in the construction of
the RSOS models \cite{abf} but, somewhat surprisingly, 
a complete analysis of the additional
symmetries of (\ref{hxxz}) when the root of unity condition
(\ref{root}) holds has never been given. 

We have found that when the root of unity condition (\ref{root}) holds 
that the Hamiltonian of the XXZ chain with periodic boundary
conditions commutes with the generators of the $sl_2$ loop algebra and
thus the space of eigenvectors decomposes into a direct sum of
finite dimensional representations of loop $sl_2.$  Moreover all of
these finite dimensional representations are made up either from
singlets or spin 1/2 representations and the dimensions of the
degenerate subspaces are all powers of 2. This algebra is very closely
related with the algebra originally used by Onsager \cite{ons} to 
solve the Ising model.

In this papers we will derive this  loop $sl_2$ symmetry at
roots of unity (\ref{root}) and use the symmetry to study the
degeneracies of the eigenvalues both of the transfer matrix of the six
vertex model and the Hamiltonian of the XXZ spin chain. 
The case $N=2$  ($\Delta=0$)
has also been solved long ago \cite{lsm} by the technique of
the Jordan Wigner transformation. From this solution it is explicitly
seen that there are many degeneracies in the spectrum which we review
in sec.2. In section 3 we present the $sl_2$ loop algebra which
generalizes the degeneracy of sec. 2 to arbitrary $N.$ 
In sec. 4 we treat the symmetry algebra for $N=2$ by means of the
Jordan--Wigner techniques of sec. 2. 
We conclude in section 5 with a discussion of how the
representation theory of loop $sl_2$ determines the degeneracy of
(\ref{hxxz}) at roots of unity and discuss the difference between
the Jordan--Wigner and the Bethe's ansatz solution for the $\Delta=0$ problem.

\section{The Jordan--Wigner solution of Lieb, Schultz and Mattis}

In 1961 Lieb, Schultz and Mattis \cite{lsm} computed the eigenvalues
of the XY Hamiltonian 
\begin{equation}
H_{XY}={1\over
2}\sum_{j=1}^L((1+\gamma)\sigma_j^x\sigma_{j+1}^x+
(1-\gamma)\sigma^y_j\sigma^y_{j+1})
\label{xy}
\end{equation}
by use of operator methods and the Jordan--Wigner transformation \cite{jw}
which reduces the Hamiltonian (\ref{xy}) involving 
Pauli spin matrices $\sigma_j^i$ ($i=x,y,z$) to a quadratic form in
anticommuting fermionic operators. When $\gamma=0$ the XY Hamiltonian  
reduces to (\ref{hxxz}) with $\Delta=0.$ In this case the total z
component of the spin
\begin{equation}
S^z={1\over 2}\sum_{j=1}^L\sigma^z_j
\label{sz}
\end{equation}
commutes with $H,$ and in the basis where $\sigma^z_j$ is diagonal 
(specified by the notation $|\pm 1 \rangle_j$ with 
$ \sigma_j^z | \pm 1 \rangle_j = \pm |\pm 1 \rangle_j$ )
the number of down spins $n$ is related to $S^z$ by 
\begin{equation}
n={L\over 2}-S^z.
\label{down}
\end{equation}
The energy eigenvalues for a given value of
$S^z\geq0$ are
\begin{equation}
E=\sum_{i=1}^{n}2\cos p_i,
\label{en0}
\end{equation}
and the corresponding momenta are
\begin{equation}
P=\sum_{j=1}^n p_j~~~({\rm mod}~~2\pi)
\label{momentum}
\end{equation}
where the $p_j$ obey the exclusion principle for momenta of free
fermions
\begin{equation}
p_i \neq p_j~~{\rm for}~~i\neq j
\label{exclusion}
\end{equation}
and are freely chosen from
\begin{equation}
p_i\in \cases{{2\pi m\over L},~m=0,1,\cdots
L-1&for $n$ odd\cr
{\pi\over L}(2m+1),~m=0,1,\cdots,L-1&for $n$ even\cr}
\label{jwmom}
\end{equation}
We also  note the reflection symmetry
\begin{equation}
E(S^z)=E(-S^z)~~~{\rm for}~S^z\neq 0.
\label{szref}
\end{equation}

The eigenvalue spectrum given by (\ref{en0})-(\ref{jwmom}) has an
important symmetry. Consider two $p_j$ such that
\begin{equation}
p_1+p_2=\pi ~~({\rm mod}~2\pi).
\label{pair}
\end{equation}
Then
\begin{equation}
\cos p_1 + \cos p_2 =0
\label{null}
\end{equation}
and thus the energy of a state with given $S^z$ is degenerate with the
state with $S^z-2$ obtained by adding the pair (\ref{pair}).

Pairs satisfying (\ref{pair}) exist for all $n$ both even and odd and
the multiplicity of the degeneracy may be computed from
(\ref{jwmom}). For definiteness we consider $L/2$ to be an integer.  
Consider first $S^z$  even. Then if $L/2$ is even (odd) $p_j$ are
obtained from the second (first) case in (\ref{jwmom}), the values
$\pi /2$ and $3\pi/2$ do not occur 
and there are $L/2$ possible pairs which satisfy (\ref{pair}). Call
$S^z_{max}$ the value of $S^z$ for which there are no pairs
satisfying (\ref{pair}). This state has $L/2-S^z_{max}$ unpaired $p_j$
and therefore the number of possible pairs which can be formed is
\begin{equation}
{L\over 2}-({L\over 2}-S^z_{max})=S^z_{max}.
\end{equation}
Thus  if we add $l$ pairs which do satisfy
(\ref{pair}) to the state with $S^z_{max}$ we obtain a state with 
\begin{equation}
S^z=S^z_{max}-2l~~~0\leq l \leq S^z_{max}
\label{szval}
\end{equation} 
and a degeneracy of
\begin{equation}
{S^z_{max}\atopwithdelims() l}.
\label{degeven}
\end{equation}

If $S^z$ is odd  and $L/2$ is even (odd) then the $p_j$ are obtained from the
first (second) case in (\ref{jwmom}). Now the values $p=\pi/2$  and $3 \pi/2$
can occur. These special values give zero contribution to the energy but cannot
participate in a pair of the form (\ref{pair}). Thus in this case
there are ${L\over 2} -1$ possible pairs. Consider first the case
where the
state $S^z_{max}$ contains either $p=\pi/2$ or $3\pi/2$ but not both. 
Then there are 
\begin{equation}
{L\over 2}-1-({L\over 2}-S^z_{max}-1)=S^z_{max}
\end{equation}
possible pairs and thus if $l$ pairs are added to the state we see that
(\ref{szval}) and (\ref{degeven}) continue to hold.

Secondly consider the case where 
$S^z_{max}$ does not contain either $p=\pi/2$ or $3 \pi/2.$
Then there are only
\begin{equation}
{L\over 2}-1-({L\over 2}-S^z_{max})=S^z_{max}-1
\end{equation}
number of possible pairs and thus if we add $l$ pairs we obtain a
state with
\begin{equation}
S^z=S^z_{max}-2l~~~0\leq l \leq S^z_{max}-1
\label{szvalodd}
\end{equation} 
and a degeneracy of
\begin{equation}
{S^z_{max}-1\atopwithdelims() l}.
\label{degodd}
\end{equation}

Finally let $S^z_{max}$ contain both $\pi/2$ and $3\pi/2.$ Then the
number of possible pairs is
\begin{equation}
{L\over 2}-1-({L\over 2}-S^z_{max}-2)=S^z_{max}+1
\end{equation}
and if we add $l$ pairs to this sate we obtain the state
\begin{equation}
S^z=S^z_{max}-2l~~~0\leq l \leq S^z_{max}+1
\label{szvalodda}
\end{equation} 
and a degeneracy of
\begin{equation}
{S^z_{max}+1\atopwithdelims() l}.
\label{degodda}
\end{equation}
These last two cases are equivalent by use of the reflection symmetry
(\ref{szref}).

\section{The symmetry of the loop  algebra $sl_2$ for general $N.$ }

Degeneracies of eigenvalues are caused by symmetries of the
system. The XY model solved by LSM \cite{lsm} is equivalent to a free
Fermi problem and this is a very powerful symmetry which leads not
only to the degeneracies of sec. 2 but to many other degeneracies as
well.

In an XXZ chain (\ref{hxxz}) of finite length $L$ the free particle
degeneracies are destroyed by letting $\Delta \neq 0.$ However the
degeneracies of section 2 are special in that they exist in
systems of finite length when the root of unity condition (\ref{root}) 
holds for $q.$ We have studied these degeneracies for chains up to
length $L=16$ for $N=3,4$ and 5 and found that there are sets of
degenerate eigenvalues for values of $S^z$ satisfying the
generalization of (\ref{szval}) to arbitrary $N$
\begin{equation}
S^z=S_{\rm max}^z-Nl.
\label{nszval}
\end{equation}
If 
\begin{equation}
S^z\equiv 0 ({\rm mod}~N)
\label{mod0}
\end{equation}
then
\begin{equation}
0\leq l \leq 2S^z_{max}/N
\end{equation}
and the degeneracy is
\begin{equation}
{2 S^z_{max}/N\atopwithdelims() l}
\end{equation}
which reduces to  (\ref{degeven}) when $N=2.$
If
\begin{equation}
S^z\neq 0 (\rm {mod}~N)
\end{equation}
then just as in section 2 there are three cases
\begin{equation}
0\leq l \leq [2S^z_{max}/N]\pm 1~~{\rm and}~~0\leq l \leq [2S^z_{\max}/N]
\end{equation}
which can occur and the degeneracies are
\begin{equation}
{[2S^z_{\max}/N]\pm 1 \atopwithdelims() l}~~{\rm and}~~
{[2 S^z_{\max}/N]\atopwithdelims() l}
\end{equation}
respectively where $[x]$ is the greatest integer contained in $x.$

These degeneracies must also be produced by a symmetry of the model
and, of course, the XXZ chain is known to follow from the commuting
transfer matrix symmetry algebra of the six vertex model. But this
holds for all values of $q$ whereas the degeneracies we are here
discussing exist only when $q$ satisfies the root of unity condition
(\ref{root}). Thus we conclude that there must be a further symmetry
beyond quantum group symmetry of the transfer matrix at generic $q.$ 

We have found that this symmetry algebra is the algebra of the loop
group $sl_2.$ In this section we will define the algebra and show how
it is related to the XXZ and six vertex model. 

The study of the symmetries of the XXZ chain with periodic boundary
conditions at roots of unity (\ref{root}) was initiated in \cite{ps} where is
was seen that even though the Hamiltonian (\ref{hxxz}) is not 
invariant under the full quantum group
$U_q(sl_2)$ it is invariant under a suitable smaller set of
operators. 
To be
specific we recall the spin 1/2 representation of the
generators of of $U_q(sl_2)$
(following ref.\cite{jimbob})
\begin{equation}
q^{S^z}=q^{\sigma^z/2}\otimes \cdots \otimes q^{\sigma^z/2}
\end{equation}
\begin{equation}
S^{\pm}=\sum_{j=1}^LS_j^{\pm}=\sum_{j=1}^Lq^{\sigma^z/2}\otimes
\cdots q^{\sigma^z/2}\otimes\sigma_j^{\pm}\otimes q^{-\sigma^z/2}\otimes
\cdots \otimes q^{-\sigma^z/2}
\label{spm}
\end{equation}
which satisfy the relations of $U_q(sl_2)$
\begin{eqnarray}
q^{S^z}S^{\pm}q^{-S^z}&=&q^{\pm 1}S^{\pm },\label{uqone}\\
{[}S^+,S^-{]}&=&{q^{2S^z}-q^{-2S^z}\over q-q^{-1}}\label{uqtwo}
\end{eqnarray}
and by definition $(a_1\otimes a_2)(b_1 \otimes b_2)=a_1b_1\otimes a_2b_2.$

The $n^{th}$ power of the operators $S^{\pm}$ satisfy
\begin{eqnarray}
S^{\pm n}&=&q^{n(n-1)/2}[n]! \sum_{1\leq j_1<j_2<\cdots<j_N\leq L}
S_{j_1}^{\pm} S_{j_2}^{\pm}\cdots S_{j_N}^{\pm}\nonumber \\
  & =&   [n]! \, \sum_{1 \le j_1 < \cdots < j_n \le L}
q^{{n \over 2 } \sigma^z} \otimes \cdots \otimes q^{{n \over 2} \sigma^z}
\otimes \sigma_{j_1}^{\pm} \otimes
q^{{(n-2) \over 2} \sigma^z} \otimes  \cdots \otimes q^{{(n-2) \over 2}
\sigma^z}
\nonumber \\
 & & \otimes \sigma_{j_2}^{\pm} \otimes q^{{(n-4) \over 2} \sigma^z} \otimes
\cdots
\otimes \sigma^{\pm}_{j_n} \otimes q^{-{n \over 2} \sigma^z} \otimes \cdots
\otimes q^{-{n \over 2} \sigma^z}
\label{npsp}
\end{eqnarray}
where 
\begin{eqnarray}
[n]&=&(q^n-q^{-n})/(q-q^{-1}) ~~{\rm for}~~n>0~~{\rm
and}~~[0]=1\nonumber\\
{[}n{]}!&=&\prod_{k=1}^n {[}k{]}
\end{eqnarray}
and we have used
\begin{equation}
q^{A\sigma^z}\sigma^{\pm}=q^{\pm A}\sigma^{\pm}.
\label{spz}
\end{equation}
Then when the root of unity condition
(\ref{root}) holds we have
\begin{equation}
(S^{\pm})^N=0.
\label{nil}
\end{equation}
However if we set 
\cite{lusa}-\cite{kac} 
\begin{equation}
S^{\pm(N)}={\rm lim}_{q^{2N}\rightarrow 1}(S^{\pm})^N/[N]!
\label{abssn}
\end{equation}
then the operators $S^{\pm(N)}$ exist and are non-vanishing
and in particular
\begin{eqnarray}
S^{\pm(N)}&=&  
\sum_{1 \le j_1 < \cdots < j_N \le L}
q^{{N \over 2 } \sigma^z} \otimes \cdots \otimes q^{{N \over 2} \sigma^z}
\otimes \sigma_{j_1}^{\pm} \otimes
q^{{(N-2) \over 2} \sigma^z} \otimes  \cdots \otimes q^{{(N-2) \over 2}
\sigma^z}
\nonumber \\
 & & \otimes \sigma_{j_2}^{\pm} \otimes q^{{(N-4) \over 2} \sigma^z} \otimes
\cdots
\otimes \sigma^{\pm}_{j_N} \otimes q^{-{N \over 2} \sigma^z} \otimes \cdots
\otimes q^{-{N \over 2} \sigma^z}
\label{sn}
\end{eqnarray}
 It is shown in \cite{ps} that for $S^z/N$ an
integer that $S^{\pm(N)}$ commutes with the Hamiltonian (\ref{hxxz})
when (\ref{root}) holds.

We have found that $S^{\pm(N)}$ is not just an isolated operator which
commutes with the Hamiltonian (\ref{hxxz}) and transfer matrix
of the six vertex model but in fact 
is part of a much larger symmetry algebra. The
key to this symmetry algebra is the observation of Jimbo \cite{jimbob} 
that the representation of $S^{\pm}$ given by (\ref{spm}) is not
unique but that there exists an equally good isomorphic representation 
\begin{equation}
T^{\pm}=\sum_{j=1}^LT_j^{\pm}=\sum_{j=1}^Lq^{-\sigma^z/2}\otimes
\cdots q^{-\sigma^z/2}\otimes\sigma_j^{\pm}\otimes q^{\sigma^z/2}\otimes
\cdots \otimes q^{\sigma^z/2}
\label{tpm}
\end{equation}
which is obtained from $S^{\pm}$ by the replacement $q\rightarrow
q^{-1}.$ In the case considered here where $q$ is a root of unity (\ref{root})
$T^{\pm}$  and $S^{\pm}$ are related by complex and hermitian conjugation
\begin{equation}
T^{\pm}=S^{\pm*}=S^{\mp\dagger}
\label{dual}
\end{equation}
which may be thought of as a dual transformation
and by
\begin{equation}
T^{\pm}=RS^{\mp}R^{-1}
\end{equation}
where $R=\sigma^x_1\otimes\sigma^x_1\otimes\cdots\otimes\sigma^x_L$ is
the spin inversion operator.
The operator  $T^{\pm}$ also has the property that when the root of
unity condition (\ref{root}) holds then $(T^{\pm})^N=0$ and we define
$T^{\pm(N)}$ exactly as we defined $S^{\pm(N)}$ by (\ref{abssn}) and 
(\ref{sn}) as
\begin{eqnarray}
T^{\pm(N)}&=&  
\sum_{1 \le j_1 < \cdots < j_N \le L}
q^{-{N \over 2 } \sigma^z} \otimes \cdots \otimes q^{-{N \over 2} \sigma^z}
\otimes \sigma_{j_1}^{\pm} \otimes
q^{-{(N-2) \over 2} \sigma^z} \otimes  \cdots \otimes q^{-{(N-2) \over 2}
\sigma^z}
\nonumber \\
 & & \otimes \sigma_{j_2}^{\pm} \otimes q^{-{(N-4) \over 2} \sigma^z} \otimes
\cdots
\otimes \sigma^{\pm}_{j_N} \otimes q^{{N \over 2} \sigma^z} \otimes \cdots
\otimes q^{{N \over 2} \sigma^z}
\label{tn}
\end{eqnarray}

By use of the theory of quantum groups it follows from the work of
Jimbo\cite{jimbob} that the elementary commutation relations 
between $S^{\pm}$ and $T^{\pm}$ 
hold for
all $q$
\begin{equation}
[S^{+},T^{+}]=[S^{-},T^{-}]=0
\label{zero}
\end{equation}
as well as the four quantum Serre relations
\begin{eqnarray}
{3 \atopwithdelims[] 0}S^{+3}T^{-}-
{3  \atopwithdelims[] 1}S^{+2}T^{-}S^{+}+
{3  \atopwithdelims[] 2}S^{+}T^{-}S^{+2}-
{3  \atopwithdelims[] 3}T^{-}S^{+3}&=&0\label{qserreone}\\
{3  \atopwithdelims[] 0}S^{-3}T^{+}-
{3  \atopwithdelims[] 1}S^{-2}T^{+}S^{-}+
{3  \atopwithdelims[] 2}S^{-}T^{+}S^{-2}-
{3  \atopwithdelims[] 3}T^{+}S^{-3}&=&0\label{qserretwo}\\
{3  \atopwithdelims[] 0}T^{+3}S^{-}-
{3  \atopwithdelims[] 1}T^{+2}S^{-}T^{+}+
{3  \atopwithdelims[] 2}T^{+}S^{-}T^{+2}-
{3  \atopwithdelims[] 3}S^{-}T^{+3}&=&0\label{qserrethree}\\
{3  \atopwithdelims[] 0}T^{-3}S^{+}-
{3  \atopwithdelims[] 1}T^{-2}S^{+}T^{-}+
{3  \atopwithdelims[] 2}T^{-}S^{+}T^{-2}-
{3  \atopwithdelims[] 3}S^{+}T^{-3}&=&0.\label{qserrefour}\\
\end{eqnarray}
and we define 
\begin{eqnarray}
{m\atopwithdelims[] l}={[m]!\over
[l]![m-l]!}&=&\prod_{k=1}^l{q^{m-l+k}-q^{-(m-l+k)}\over q^k-q^{-k}}
~~{\rm for}~0<l<m\nonumber\\
&=&1~~{\rm for}~l=0,m
\label{qbin}
\end{eqnarray}

We will give elementary proofs of these fundamental commutation relations
which do not explicitly rely on quantum group theory later in this section.

From these equations we specialize to $q^{2N}=1$ to derive
\begin{equation}
[S^{+(N)},T^{+(N)}]=[S^{-(N)},T^{-(N)}]=0
\label{one}
\end{equation}

\begin{equation}
[S^{\pm(N)},S^z]=\pm NS^{\pm(N)},~~~[T^{\pm(N)},S^z]=\pm N T^{\pm(N)}
\label{three}
\end{equation}
and, by use of the higher order Serre relations of Lusztig
\cite{lusc}, we will show that 
\begin{eqnarray}
S^{+(N)3}T^{-(N)}-3S^{+(N)2}T^{-(N)}S^{+(N)}+3S^{+(N)}T^{-(N)}S^{+(N)2}
-T^{-(N)}S^{+(N)3}&=&0 \label{serreone}\\
S^{-(N)3}T^{+(N)}-3S^{-(N)2}T^{+(N)}S^{-(N)}+3S^{-(N)}T^{+(N)}S^{-(N)2}
-T^{+(N)}S^{-(N)3}&=&0 \label{serretwo}\\
T^{+(N)3}S^{-(N)}-3T^{+(N)2}S^{-(N)}T^{+(N)}+3T^{+(N)}S^{-(N)}T^{+(N)2}
-S^{-(N)}T^{+(N)3}&=&0 \label{serrethree}\\
T^{-(N)3}S^{+(N)}-3T^{-(N)2}S^{+(N)}T^{-(N)}+3T^{-(N)}S^{+(N)}T^{-(N)2}
-S^{+(N)}T^{-(N)3}&=&0 
\label{serrefour}
\end{eqnarray}
In the sector $S^z\equiv 0 ({\rm mod}~N)$ we additionally by  use of
results from \cite{kactwo} that
\begin{equation}
[S^{+(N)},S^{-(N)}]=[T^{+(N)},T^{-(N)}]=-(-q)^N{2\over N}S^z.
\label{two}
\end{equation}
Proofs will be given later in this section.

If we make the  identifications
\begin{equation}
e_0=S^{+(N)},~~f_0=S^{-(N)},~~e_1=T^{-(N)},~~f_1=T^{+(N)},t_0=-t_1=-(-q)^NS^z/N
\end{equation}
and if we do not impose
the relation (\ref{dual}) and the identity $t_0=-t_1$ which was demanded by
(\ref{two}) we see that equations (\ref{one})-(\ref{two}) 
are the defining
relations \cite{kactwo} of the Chevalley generators of the 
affine Lie algebra $A^{(1)'}_1.$
The identity $t_0=-t_1$ reduces this algebra  to the defining
relations of the Chevalley generators of the loop algebra of $sl_2.$

The theory of $sl_2$ loop algebras and of its finite dimensional 
representations may now be
applied to the XXZ model when $q$ satisfies the root of unity
condition (\ref{root}) by noting the commutation relation with the
Hamiltonian (\ref{hxxz}) which holds when $S^z\equiv 0 (\rm{mod}~N)$
\begin{eqnarray}
{[}S^{\pm(N)},H{]}={[}T^{\pm(N)},H{]}=0.
\label{sthcomm}
\end{eqnarray}
More generally we have translational (anti)invariance of the operators
$S^{\pm (N)},~T^{\pm (N)}$
\begin{equation}
S^{\pm (N)}e^{iP}=q^Ne^{iP}S^{\pm (N)},
~~~T^{\pm (N)}e^{iP}=q^Ne^{iP}T^{\pm (N)}
\label{trans}
\end{equation}
where the momentum operator $P$ is defined from the shift operator
\begin{equation}
\Pi_L|_{\{j\},\{j'\}}=\prod_{i=1}^L\delta_{j_i,j'_{i-1}}
\label{shift}
\end{equation}
as $\Pi_L=e^{-iP}$ 
and we will prove in appendix A the (anti) commutation relation 
\begin{equation}
S^{\pm (N)}T(v)=q^NT(v)S^{\pm (N)},~~~T^{\pm (N)}T(v)=q^NT(v)T^{\pm (N)} 
\label{stcomm}
\end{equation}
where $T(v)$ is the six vertex model
transfer matrix. Thus
\begin{equation}
{[}S^{\pm(N)},T(v)e^{-iP}{]}={[}T^{\pm(N)},T(v)e^{-iP}{]}=0
\label{sixvc}
\end{equation}
which reduces to (\ref{sthcomm}) as 
$e^v \rightarrow 1.$
Therefore the spectrum of the transfer matrix of the six vertex
model and the spectrum of the XXZ model decomposes into finite
dimensional  
representations of the $sl_2$ loop algebra which are given explicitly
on page 243 of \cite{kactwo}

The discussion of the previous section indicates that we should expect that
for sectors where $S^z\neq 0 ({\rm mod}~N)$ 
the existence of an $sl_2$ loop algebra is
somewhat more involved. We begin by noting that in the sector
$S^z\equiv n ({\rm mod}~N)$ where $1\leq n \leq N-1$ 
that the following 
four operators are translationally invariant and commute 
with the transfer matrix
\begin{equation}
  (T^{+})^n(S^{-})^n,~~ (S^{+})^n(T^{-})^n,~~ (T^{-})^{N-n}(S^{+})^{N-n},~~
 (S^{-})^{N-n}(T^{+})^{N-n} 
\label{proj}
\end{equation}
Furthermore even though $S^{\pm(N)}$ and $T^{\pm(N)}$ do not (anti) commute
with the transfer matrix that the eight  operators
\begin{eqnarray}
&~&(T^{+})^n(S^{-})^nS^{-(N)},~~
S^{-(N)} (S^{-})^{N-n}(T^{+})^{N-n}\nonumber \\
&~&S^{+(N)} (S^{+})^n(T^{-})^n,~~
(T^{-})^{N-n}(S^{+})^{N-n}S^{+(N)},\nonumber \\
&~&T^{+(N)}(T^{+})^n(S^{-})^n,~~ 
(S^{-})^{N-n}(T^{+})^{N-n}T^{+(N)}\nonumber \\
&~&(S^{+})^n(T^{-})^nT^{-(N)},~~ 
T^{-(N)}(T^{-})^{N-n}(S^{+})^{N-n},
\label{Nproj}
\end{eqnarray}
do (anti) commute both with $e^{iP}$ and the transfer matrix
in the sector $S^z\equiv n ({\rm mod}~N).$

The operators of (\ref{proj}) each have a large null space
but they are not themselves projection operators. However on the
computer (at least for $N=3$) we have numerically constructed the
projection operators onto the eigenspace of the nonzero eigenvalues
of the operators(\ref{proj}) and using
(\ref{Nproj}) have constructed the corresponding projections
 of
$S^{\pm(N)}$ and $T^{\pm(N)}$ and have verified that 
for the projected operators in the sector $S^z\equiv
n ({\rm mod}N)$ with $1\leq n \leq N-1$ we find that
(\ref{one})-(\ref{serrefour}) hold
without modification but (\ref{two}) is slightly modified.

We also note that the case of $N=2$ is special
in that the operator $T^{+}S^{-}$ satisfies 
\begin{equation}
(T^{+}S^{-})^2=LT^{+}S^{-}
\label{n2proj}
\end{equation}
for all values of $S^z$ (and not just $S^z\equiv 1 (\rm{mod}~2)$) 
and similarly for the other three operators  
$S^{+}T^{-},~T^{-}S^{+},~S^{-}T^{+}$.
Thus (up to a factor of $L$) these operators for $N=2$ are
 already projection operators.

We conclude this section with the proofs of (\ref{zero})-
(\ref{two}) and (\ref{trans}).

\subsection{Proof of (\ref{zero}) and (\ref{one}).}

To prove (\ref{zero}) we use the definitions (\ref{spm}) and
(\ref{tpm}) and the fact that $(\sigma^+)^2=0$ to write
\begin{eqnarray}
[S^{+},T^{+}]&=&\sum_{j_1<j_2}\{(q^{\sigma^z/2}\sigma_{j_1}^+)\otimes
q^{\sigma^z}\otimes \cdots\otimes
q^{\sigma^z}\otimes(\sigma_{j_2}^+q^{\sigma^z/2})\nonumber\\
&+&(\sigma_{j_1}^+q^{-\sigma^z/2})\otimes
q^{-\sigma^z}\otimes \cdots\otimes
q^{-\sigma^z}\otimes (q^{-\sigma^z/2}\sigma_{j_2}^+)\nonumber \\
&-& (q^{-\sigma^z/2}\sigma_{j_1}^+)\otimes
q^{-\sigma^z}\otimes \cdots\otimes
q^{-\sigma^z}\otimes (\sigma_{j_2}^+q^{-\sigma^z/2})\nonumber \\
&-&(\sigma_{j_1}^+q^{\sigma^z/2})\otimes
q^{\sigma^z}\otimes \cdots\otimes
q^{\sigma^z}\otimes (q^{\sigma^z/2}\sigma_{j_2}^+)\}.
\end{eqnarray}
If we now use the commutation relation(\ref{uqone}) it is seen that
terms 1 and 4 and terms 2 and 3 cancel in this sum. Thus(\ref{zero})
follows.
Equation (\ref{one}) follows immediately from (\ref{zero}).

\subsection{Proof of (\ref{qserreone})-(\ref{qserrefour})}

To give an elementary proof of (\ref{qserreone}) we divide by $[3]!$ and
use both (\ref{npsp}) and the companion equation for $T^{\pm n}$ to write
\begin{eqnarray}
& &(S^{+3}/[3]!)T^{-}-(S^{+2}/[2]!)T^-S^{-}+S^{+}T^{-}(S^{-2}/[2]!)-T^
-(S^{+3}/[3]!)\nonumber\\
&=&(\sum_{1\leq j_1<j_2<j_3\leq L}q^{{3\over 2}\sigma^z}
\otimes\sigma_{j_1}^+\otimes
q^{{1\over 2}\sigma^z}\otimes\sigma_{j_2}^+\otimes
q^{-{1\over 2}\sigma^z}\otimes\sigma_{j_3}^+\otimes q^{-{3\over
2}\sigma^z})(\sum_{1\leq j_4\leq L}q^{-{1\over
2}\sigma^z}\otimes\sigma^-_{j_4}\otimes q^{{1\over
2}\sigma^z})\nonumber\\
&-&(\sum_{1\leq j_1<j_2\leq L}q^{\sigma^z}\otimes
\sigma_{j_1}^+\otimes I \otimes \sigma_{j_2}^+\otimes q^{-{\sigma^z}})
(\sum_{1\leq j_4 \leq L}q^{-{1\over 2}\sigma^z}\otimes \sigma_{j_4}^-\otimes
q^{{1\over 2}\sigma^z})
(\sum_{1\leq j_3 \leq L}q^{{1\over 2}\sigma^z}\otimes \sigma_{j_3}^+\otimes
q^{-{1\over 2}\sigma^z})\nonumber\\
&+&(\sum_{1\leq j_1 \leq L}q^{{1\over 2}\sigma^z}\otimes \sigma_{j_1}^+\otimes
q^{-{1\over 2}\sigma^z})
(\sum_{1\leq j_4 \leq L}q^{-{1\over 2}\sigma^z}\otimes \sigma_{j_4}^-\otimes
q^{{1\over 2}\sigma^z})
(\sum_{1\leq j_2<j_3\leq L}q^{\sigma^z}\otimes
\sigma_{j_2}^+\otimes I \otimes \sigma_{j_3}^+
\otimes q^{-{\sigma^z}})\nonumber\\
&-&(\sum_{1\leq j_4\leq L}q^{-{1\over
2}\sigma^z}\otimes\sigma^-_{j_4}\otimes q^{{1\over
2}\sigma^z})
(\sum_{1\leq j_1<j_2<j_3\leq L}q^{{3\over 2}\sigma^z}
\otimes\sigma_{j_1}^+\otimes
q^{{1\over 2}\sigma^z}\otimes\sigma_{j_2}^+\otimes
q^{-{1\over 2}\sigma^z}\otimes\sigma_{j_3}^+\otimes q^{-{3\over
2}\sigma^z})
\label{serrehelp}
\end{eqnarray}
where we have used $q^{a\sigma^z}$ to
denote $q^{a\sigma^z}\otimes\cdots \otimes q^{a\sigma^z}.$ 

We now note that in the expansion of this expression there are two
types of terms, those where $\sigma^-_{j_4}$ never is at the same site
as one of the $\sigma^+_j$ and those where $\sigma^-_{j_4}$ and at
least  one of
the $\sigma^+_j$ are at the same site. These two types are treated
separately. 

For the first type of term there are four distinct cases depending on
the location of the $\sigma^-_{j_4}$ relative to the three $\sigma^+_j.$
For example consider the term where $\sigma^-_{j_4}$ lies to the right
of the three $\sigma^+_j.$ Then each of the four terms in
(\ref{serrehelp}) gives a contribution of the form

\begin{equation}
A_j\sum_{1\leq j_1<j_2<j_3<j_4\leq L}q^{\sigma^z}\sigma^+_{j_1}\otimes
I \otimes \sigma^+_{j_2}\otimes q^{-\sigma^z}\otimes
\sigma^+_{j_3}\otimes q^{-2\sigma^z}\otimes \sigma^-_{j_4}\otimes
q^{-\sigma^z}
\label{serrehelp2}
\end{equation}
where by use of (\ref{spz}) it is elementary  to find
\begin{equation}
A_1=q^3,~~A_2=-q^3-q-q^{-1},~~A_3=q+q^{-1}+q^{-3},~~A_4=-q^{-3}.
\end{equation}
We see that $A_1+A_2+A_3+A_4=0$ and thus the contribution from
these terms vanished.

Similar elementary computations demonstrate similar cancelations in
all other cases and thus (\ref{qserreone}) is demonstrated. The other
Serre relations (\ref{qserretwo})--(\ref{qserrefour}) follow in the
identical manner.

\subsection{Proof of (\ref{two}).}

We begin the proof of (\ref{two}) by noting the commutation relation
valid for general $q$ (1.3.1) on page 474 of \cite{kac} for the case
of the group $U_q(sl_2)$
\begin{equation}
[(S^{+})^m,(S^{-})^n]=\sum_{j=1}^{\rm{min}(m,n)}{m\atopwithdelims[] j} 
{n\atopwithdelims[] j}[j]!(S^{-})^{n-j}(S^{+})^{m-j}\prod_{k=0}^{j-1}
{q^{2S^z+m-n-k}-q^{-(2S^z+m-n-k)}\over {q-q^{-1}}}.
\end{equation}
Thus setting $m=n=N$ and dividing by $[N]!^2$ we obtain
\begin{equation}
[(S^{+})^N/[N]!,(S^{-})^N/[N]!]=\sum_{j=1}^{N}{1\over
  [N-j]!^2[j]!}(S^{-})^{N-j}
(S^{+})^{N-j}\prod_{k=0}^{j-1}{q^{2S^z-k}-q^{-2S^z+k}\over q-q^{-1}}.
\end{equation}
In the sector $S^z\equiv 0~{\rm mod}~ N$ the $k=0$ term in the product above
vanishes in the limit when $q^{2N}\rightarrow 1.$ Thus the only terms in the sum
over $j$ which can fail to vanish are those where the coefficient of
the product diverge. This occurs only for $j=N.$ Thus 
using the
definition (\ref{abssn}) we may  let $q^{2N}\rightarrow 1$ to find
\begin{eqnarray}
[S^{+(N)},S^{-(N)}]&=&{\rm lim}_{q^{2N}\rightarrow 1}{1\over
  [N]!}\prod_{k=0}^{N-1}{q^{2S^z-k}-q^{-2S^z+k}\over q-q^{-1}}\\
&=&{(-1)^N\over[N]}{\rm lim}_{q^{2N} \rightarrow 1}{q^{2S^z}-q^{-2S^z}\over
  q^N-q^{-N}}\\
&=&-(-q)^N{2\over N}S^z
\end{eqnarray}
as desired.

\subsection{Proof of (\ref{serreone})-(\ref{serrefour})}

To proof (\ref{serreone})-(\ref{serrefour}) we begin with 
a result of Lusztig ((7.1.6)
on page 57 of ref. \cite{lusc}) on higher order q-Serre relations. That result 
when specialized to the quantum group $U_q(\widehat{sl_2})$ says that for the
operators $S^+$ and $T^-$ which satisfy the quantum Serre relation 
(\ref{qserreone})
it follows from properties of the algebra alone (and not of the representation)
the for any $q$ if we set
\begin{eqnarray}
\theta_1^{(m)}&=&(S^+)^m/[m]!\nonumber\\
\theta_2^{(m)}&=&(T^-)^m/[m]!
\label{thetadfn}
\end{eqnarray}
then we have
\begin{equation}
\theta_1^{(3N)}\theta_2^{(N)}=\sum_{s'=N}^{3N}\gamma_s'
\theta_1^{(3N-s')}\theta_2^{(N)}\theta_1^{(s')}
\label{lustone}
\end{equation}
where 
\begin{equation}
\gamma_s'=(-1)^{s'+1}q^{s'(N-1)}\sum_{l=0}^{N-1}(-1)^lq^{l(1-s')}
{s'\atopwithdelims[] l}.
\label{lustwo}
\end{equation}

We obtain the Serre relation (\ref{serreone}) taking the limit $q^{2N}
\rightarrow 1$ in
(\ref{lustone}).
Thus we write
\begin{equation}
s'=Ns+p~~~{\rm with}~~p=0,1,\cdots N-1.
\end{equation}
and by use of (\ref{qbin}) we see that
\begin{equation}
{\rm lim}_{q^{2N}\rightarrow
1}{sN+p\atopwithdelims[] l}=\cases{q^{Nsl}{p\atopwithdelims[] l}&for
$p\geq l$\cr
0&otherwise}
\end{equation}
Therefore we obtain
\begin{equation}
{\rm lim}_{q^{2N}\rightarrow 1}\gamma_{Ns+p}=(-1)^{Ns+1}q^{(Ns+p)(N-1)}
\sum_{l=0}^p(-1)^lq^{l(1-p)}{p\atopwithdelims[] l}.
\end{equation}
However from the q-binomial theorem (for example 1.34 of
ref\cite{lusc}) we have
\begin{equation}
\sum_{l=0}^p(-1)^lq^{l(1-p)}{p\atopwithdelims[] l}=\delta_{p,0}
\end{equation}
for all $q$ and thus we obtain
\begin{equation}
{\rm lim}_{q^{2N}\rightarrow 1}\gamma_{Ns+p}=\delta
(-1)^{Ns+1}q^{Ns(N-1)}=\delta_{p,0}(-1)^{s+1}
\label{limgam}
\end{equation}
where in the last line we have used $q^N=\pm 1$ for $N$ odd and
$q^N=-1$ for $N$ even.

If we now note from (\ref{thetadfn}) that
\begin{equation}
\theta^{(Ns)}={[N]!^s\over [Ns]!}\theta^{(N)s}
\end{equation}
and use the relation derived from (\ref{qbin}) that
\begin{equation}
{\rm lim}_{q^{2N}\rightarrow 1}{[N]!^s\over [Ns]!}={q^{N^2}\over s!}
\end{equation}
we may use (\ref{limgam}) in (\ref{lustone}) and restore the definition
(\ref{thetadfn}) to find
\begin{equation}
\sum_{s=0}^3{(-1)^{s+1}\over s!(3-s)!}S^{+(N)3-s}T^{-(N)}S^{+(N)s}=0
\end{equation}
from which (\ref{serreone}) follows immediately. The proof of the
remaining Serre relations (\ref{serretwo})-(\ref{serrefour}) is identical.

\subsection{Proof of translational (anti)invariance of $S^{\pm (N)}$
in $S^z\equiv 0 ({\rm mod}~N)$ (\ref{trans}) and  
$(T^+)^n(S^{-})^n$ in $S^z\equiv n ({\rm mod}~N)$ (\ref{proj})}

\par  
Let us denote by $\Pi_R$ the inverse of 
the shift operator $\Pi_L$.  
By definition of the shift operator (\ref{shift}) we have for any set
of operators $A_j$ in the $j^{th}$ position in the tensor product
\begin{equation}
\Pi_R A_1\otimes A_2\otimes \cdots \otimes A_L \Pi_R^{-1} 
= A_2\otimes
A_3\otimes A_L\otimes A_1 \, . 
\label{defshift}
\end{equation}
Considering the actions of the shift 
operator $\Pi_R$ on 
$ S_j^{\pm}$ ($T_j^{\pm}$)
for $j=1, \ldots, L$, explicitly,   we have 
\begin{eqnarray}
\Pi_R S^{\pm} \Pi_R^{-1} & = & (S^{\pm} -S_L^{\pm}) q^{\sigma_L^z} 
+ S_L^{\pm} q^{-2S^z + \sigma_L^z }  \, , 
\label{RshiftS} \\
\Pi_R T^{\pm} \Pi_R^{-1} & = & (T^{\pm} -T_L^{\pm}) q^{-\sigma_L^z} 
+ T_L^{\pm} q^{2S^z - \sigma_L^z }  \, .  
\label{RshiftT}
\end{eqnarray} 
and we note that $\sigma_L^z$ commutes with 
$(S^{\pm} - S_L^{\pm})$ and $(T^{\pm} - T_L^{\pm})$ . 
Taking the $n$th powers of  (\ref{RshiftS}) and (\ref{RshiftT}), 
we find, for all $q$ 
\begin{eqnarray}
\left( \Pi_R S^{\pm} \Pi_R^{-1} \right)^n 
& = & 
\left\{ 
(S^{\pm})^n + q^{\pm(n-1)} [n] 
(S^{\pm})^{n-1}S_L^{\pm} (q^{-2S^z} -1) 
\right\} q^{n \sigma_L^z} 
\label{nSR} \\ 
& = & q^{n \sigma_L^z} \left\{ 
(S^{\pm})^n + q^{\pm(n-1)} [n] 
(S^{\pm})^{n-1}S_L^{\pm} (q^{-2(S^z \pm n)} -1) 
\right\} 
\label{nSL} \\
\left( \Pi_R T^{\pm} \Pi_R^{-1} \right)^n 
& = & \left\{ 
(T^{\pm})^n + q^{\mp(n-1)} [n] 
(T^{\pm})^{n-1}T_L^{\pm} (q^{2S^z} -1) 
\right\} q^{- n \sigma_L^z}  
\label{nTR} \\ 
& = & 
q^{- n \sigma_L^z} \left\{ 
(T^{\pm})^n + q^{\mp(n-1)} [n] 
(T^{\pm})^{n-1} T_L^{\pm} (q^{2(S^z \pm n)} -1) 
\right\} 
\label{nTL}
\end{eqnarray}

The commutation relation (\ref{trans}) for $S^z\equiv 0 ({\rm mod} N)$
now follow from (\ref{nSL}) and (\ref{nTL}) by letting $n=N,$ dividing
by $[N]!$, taking the limit $q^{2N}\rightarrow 1$ and 
using $q^{-2(S^z\pm N)}=1.$

Similarly for $(S^+)^n(T^-)^n$ we find from (\ref{nSR})-(\ref{nTL})
for arbitrary $q$
\begin{eqnarray} 
\Pi_R \, (S^+)^n (T^-)^n    
& = & 
\left( \Pi_R S^+ \Pi_R^{-1} \right)^n 
\left( \Pi_R T^- \Pi_R^{-1} \right)^n 
\, \Pi_R   \nonumber \\ 
& = & 
\left\{ 
(S^{+})^n + q^{n-1} [n]   
(S^{+})^{n-1}S_L^{+} (q^{-2S^z} -1) 
\right\} q^{n \sigma_L^z} 
\nonumber \\
& & \times 
q^{- n \sigma_L^z} \left\{ 
(T^{-})^n + q^{-(n-1)} [n] 
(T^{-})^{n-1} T_L^{-} (q^{2(S^z - n)} -1) 
\right\} 
\, \Pi_R     
\label{CR_PR} 
\end{eqnarray} 
Thus if we use $q^{S^z}(T^-)^n=(T^-)^nq^{S^z-n}$ we find
$S^z=n>0$ that for all $q$ 
\begin{equation}
[\Pi_R,(S^+)^n(T^-)^n]=0.
\end{equation}
When the root of unity condition $q^{2N}=1$ holds  
this  argument immediately extends to $S^z=n ({\rm mod}~N).$


\section{A Jordan--Wigner proof of the symmetry for $N=2$}

 In this section we will prove all of the loop $sl_2$ commutation
 relations for $N=2$ by use of the Jordan--Wigner operators used by
 LSM \cite{lsm} in the computation of the eigenvalue spectrum discussed
 in sec. 2. 
 This construction
 provides insight into the general representation theory of the loop
 algebra $sl_2$ and into the projection operators needed for $S^z\equiv 1
 ({\rm mod} 2).$

 \subsection{Notation}

 We begin by recalling the Jordan--Wigner transformation of ref. \cite{lsm}
 The Hamiltonian (\ref{hxxz}) with $\Delta=0$ 
 \begin{equation}
 H = \frac{1}{2}\sum_{j=1}^{L}(\sigma_{j}^{x}\sigma_{j+1}^{x} 
 + \sigma_{j}^{y}\sigma_{j+1}^{y})
 \label{eq.XXZq}
 \end{equation}
 is first written in terms of 
 $\sigma_j^{\pm}={1\over 2}(\sigma_j^x\pm i\sigma_j^y)$
 as
 \begin{equation}
 H = \sum_{j=1}^{L}(\sigma_{j}^{+}\sigma_{j+1}^{-}+
 \sigma_{j}^{-}\sigma_{j+1}^{+}).
 \end{equation}
 We then define the
 Jordan-Wigner transformation to operators $c_j$ and $c_j^{\dagger}$
 \begin{eqnarray}
 c_{j}&=&\exp(i\pi\sum_{k=1}^{j-1}\sigma_{k}^{+}\sigma_{k}^{-})\sigma_j^-=
 e^{{\pi i\over 2}(j-1)}
 \exp({i\pi\over 2}\sum_{k=1}^{j-1}\sigma^z_k)\sigma_j^-=
 e^{{-\pi i\over 2}(j-1)}
 \exp({-i\pi\over 2}\sum_{k=1}^{j-1}\sigma^z_k)\sigma_j^-\nonumber\\
 c^{\dagger}_{j}&=&\exp(-i\pi\sum_{k=1}^{j-1}\sigma_{k}^{+}\sigma_{k}^{-})
 \sigma_{j}^{+}=e^{-{\pi i\over 2}(j-1)}
 \exp({-i\pi\over 2}\sum_{k=1}^{j-1}\sigma^z_k)\sigma_j^+
 =e^{{\pi i\over 2}(j-1)}
 \exp({i\pi\over 2}\sum_{k=1}^{j-1}\sigma^z_k)\sigma_j^+
 \label{c}
 \end{eqnarray}
 with
 \begin{equation}
 c_j^{\dagger} c_j=\sigma^+_j\sigma^-_j={1\over 2}(1+\sigma_j^z)
 \end{equation}
 which has the inverse
 \begin{equation}
 \sigma_{j}^{-}=\exp(
 i\pi\sum_{k=1}^{j-1}c_{k}^{\dagger}c_{k})c_{j},~~~
 \sigma_{j}^{+}=\exp(-i\pi\sum_{k=1}^{j-1}c_{k}^{\dagger}c_{k})c_{i}^{\dagger}.
 \label{inverse}
 \end{equation}
 Here we have used the  fact that because the eigenvalues
 of $\sigma_k^+\sigma_k^-$ are only zero and one,  the reality condition
 holds  
 \begin{equation}
 c_j=c_j^*.
 \label{real}
 \end{equation}
 Furthermore the $c_j$ and $c_j^{\dagger}$ satisfy Fermi canonical
 anticommutation relations
 \begin{equation}
 \{c_j,c_{j'}^{\dagger}\}=
 \delta_{j,j'},~~{\rm and}~~\{c_j,c_{j'}\}=\{c^{\dagger}_j,
 c^{\dagger}_{j'}\}=0
 \label{anticomm}
 \end{equation}
 In terms of these new operators the Hamiltonian (\ref{eq.XXZq}) becomes
 \begin{equation}
 H =  \sum_{i=1}^{L-1}(c_{i}^{\dagger}c_{i+1}-c_{i}c_{i+1}^{\dagger}) 
 +(c_L^{\dagger}c_1-c_Lc_1^{\dagger})e^{\pi i(S^z+L/2-1)}
 \label{hamc}
 \end{equation}
 where the final term may be interpreted as a periodic or antiperiodic
 boundary condition by defining
 \begin{equation}
 c_{L+1}=c_1e^{\pi i (l+1)}~~{\rm with}~~l=S^z+L/2
 \end{equation}
 where $l$ is the number of up spins.

 The two different boundary conditions on (\ref{hamc}) are treated
 together by introducing
 for $l\equiv 0,1~~({\rm mod}~2)$ the  
 Fourier transform operators $\eta^{(l)}_p$ and
 $\eta_p^{(l)\dagger}$ by
 \begin{equation}
 \eta^{(l)}_{p} = \frac{1}{\sqrt L}\sum_{k=1}^{L}\exp(-i k p)c_{k},~~
 \eta_{p}^{(l)\dagger} = 
 \frac{1}{\sqrt L}\sum_{k=1}^{L}\exp(i k p)c_{k}^{\dagger}
 \label{eta}
 \end{equation}
 with the inverse
 \begin{equation}
 c_{k} = \frac{1}{\sqrt L} \sum_{p} \exp(i k p) \eta^{(l)}_{p},~~
 c_{k}^{\dagger} = \frac{1}{\sqrt L} \sum_{p} 
 \exp(-i k p) \eta_{p}^{(l)\dagger}
 \end{equation}
 where for $p$ we may use either of the 
 allowed sets of $p$ which  follow from the periodicity 
 requirements $c_{k+L} =(-1)^{(l+1)}\exp(i L p)c_{k}$  
 \begin{equation}
 p=\cases{{\pi\over L}(2m+1),~~0\leq m \leq L-1&for $l=0$ \cr
 {\pi\over L}2m,~~0\leq m \leq L-1& for $l=1$}
 \label{momentum1}
 \end{equation}
 just as in sec. 2. From the reality condition (\ref{real})
 we have
 \begin{equation}
 \eta_p^*=\eta_{2\pi-p},~~~\eta_p^{*\dagger}=\eta_{2\pi-p}^{\dagger},
 \end{equation}
 from (\ref{anticomm}) and (\ref{eta})we obtain  
 \begin{equation}
 \{\eta^{(l)}_{p},\eta_{p'}^{(l)\dagger}\} = \delta_{p,p'} ~~{\rm and}~~
 \{\eta^{(l)}_{p},\eta^{(l)}_{p'}\} = 
 \{\eta_{p'}^{(l)\dagger},\eta_{p}^{(l)\dagger}\} = 0
 \label{etacomm}
 \end{equation}
 and from (\ref{hamc}) we find
 \begin{equation}
 H_l =  2 \sum_{p}\cos(p)\eta^{(l)\dagger}_{p}\eta^{(l)}_{p}.
 \label{hldef}
 \end{equation}
 The eigenstates of $H_l$ are
 \begin{equation}
 \eta^{(l)\dagger}_{p_{1}}\eta^{(l)\dagger}_{p_{2}}\cdots
 \eta^{(l)\dagger}_{p_{n}}|0>
 \label{state}
 \end{equation}
 where $|0>$ is the state with all $L$ spins down
 and the number of fermions 
 \begin{equation}
 n=\sum_{m=1}^{L} c^{\dagger}_{m}c_{m} = \sum_{p} \eta^{\dagger}_{p}\eta_{p}=L/2+S^z
 \label{number}
 \end{equation}
 takes the values $0,1,\cdots,L$.
 The number of states (\ref{state}) is $2^{L}$.
 To obtain the coordinate representations of $H-$eigenstates we use (\ref{eta})
 and (\ref{c}).

 \subsection{The operators of the loop algebra $sl_2$}

 For $N=2$ the operators $S^{\pm}$ and $T^{\pm}$ are defined from
 (\ref{spm}) and (\ref{tpm})
 and the operators $S^{\pm (2)}$ and $T^{\pm(2)}$ are defined from
 (\ref{sn}) and (\ref{tn}) with $q=e^{\pi i/2}.$ 
 We may write them
 explicitly in terms of $c_j$ and $c_j^{\dagger}$ as
 \begin{eqnarray}
 S^+&=&e^{\pi i/4}(\sum_{j=1}^Le^{-\pi i j/2}c_j^{\dagger})e^{-\pi i S^z/2}=
 q^{3\pi i/4}e^{-\pi i S^z/ 2}\sum_{j=1}^Le^{-\pi i j/ 2}
 c_j^{\dagger}\nonumber\\
 T^+&=&e^{-\pi i / 4}(\sum_{j=1}^Le^{\pi i j/2}
 c_k^{\dagger})e^{\pi i S^z/2}=q^{-3\pi i/4}e^{\pi i
 S^z/2}\sum_{j=1}^Le^{\pi i j/2}c^{\dagger}_j
 \label{stc}
 \end{eqnarray}
 and
 \begin{eqnarray}
 S^{+(2)}&=&\sum_{1\leq j<k\leq L}
 e^{{\pi
 i\over2}\sum_{m=1}^{j-1}\sigma^z_m}\sigma_j^{+}\sigma_k^{+}e^{-{\pi
 i\over 2}\sum_{m=k+1}^L\sigma_m^z}\nonumber\\
 &=&ie^{\pi i S^z}\sum_{1\leq j<k\leq L}e^{-{\pi i\over
 2}(j+k)}c_j^{\dagger}c_k^{\dagger}
 \label{sp2c}
 \end{eqnarray}
 \begin{eqnarray}
 T^{+(2)}&=&\sum_{1\leq j<k\leq L}
 e^{-{\pi
 i\over2}\sum_{m=1}^{j-1}\sigma^z_m}\sigma_j^{+}\sigma_k^{+}e^{{\pi
 i\over 2}\sum_{m=k+1}^L\sigma_m^z}\nonumber\\
 &=&-ie^{\pi i S^z}\sum_{1\leq j<k\leq L}e^{{\pi i\over
 2}(j+k)}c_j^{\dagger}c_k^{\dagger}
 \label{tp2c}
 \end{eqnarray}
 and $S^{-}, T^{-}, S^{-(2)}$ and $T^{-(2)}$ are obtained by the replacements
 $\sigma_j^+\rightarrow\sigma_j^-$ and $c_j^{\dagger}\rightarrow c_j$ in
 (\ref{stc}),(\ref{sp2c}) and (\ref{tp2c}). 
 Here we have used the identity
 $\sigma_j^z/2=\sigma_j^+\sigma_j^- -1/2,$ and  the definition (\ref{c}).

 We may now use the Fourier transform operators (\ref{eta}) and for the
 operators $S^{\pm},~T^{\pm}$ we easily find
 \begin{eqnarray}
 S^{+}&=&L^{1/2}e^{\pi i /4}\eta_{3\pi/2}^{\dagger}e^{-\pi i S^z/2}=
 L^{1/2}e^{3 \pi i/4}e^{- \pi i S^z/2}\eta_{3\pi/2}^{\dagger}
 \nonumber\\
 T^{+}&=&L^{1/2}e^{-\pi i/4}\eta^{\dagger}_{\pi/2}e^{\pi i
 S^z/2}=L^{1/2}e^{-3\pi i/4}e^{\pi i S^z/2}\eta^{\dagger}_{\pi/2}
 \nonumber\\
 S^{-}&=&L^{1/2}e^{3\pi i/4}\eta_{\pi/2}e^{-\pi i S^z/2}
 =L^{1/2}e^{\pi i/4}e^{-\pi i S^z/2}\eta_{\pi /2}
 \nonumber\\
 T^{-}&=&L^{1/2}e^{-3\pi i/4}\eta_{3\pi/2}e^{\pi i S^z/2}=L^{1/2}e^{-\pi
 i/4}e^{\pi i S^z/2}\eta_{3\pi/2}.
 \label{stcform}
 \end{eqnarray}
 From this equation and the anticommutation relations (\ref{etacomm})
 we find that
 \begin{equation}
 [S^+,T^+]=-iL\{\eta_{3\pi/2}^{\dagger},
 \eta_{\pi/2}^{\dagger}\}=0,~~~
 [S^{-},T^{-}]=-iL\{\eta_{\pi/2},\eta_{3\pi/2}\}=0
 \end{equation}
 which is the commutation relation (\ref{zero}).
 In addition we see from (\ref{stcform}) that 
 \begin{equation}
 S^+T^-=L\eta_{3\pi\over 2}^{\dagger}\eta_{3\pi\over 2},~~
 T^-S^+=L\eta_{3\pi\over 2}\eta_{3\pi\over 2}^{\dagger},~~
 T^+S^-=L\eta_{\pi\over 2}^{\dagger}\eta_{\pi\over 2},~~
 S^-T^+=L\eta_{\pi\over 2}\eta_{\pi\over 2}^{\dagger}.
 \end{equation}
 from which we get the projection operator relations (\ref{n2proj})
 \begin{equation}
 (S^+T^-)^2=LS^+T^-,~~(T^-S^+)^2=LT^-S^+,
 ~~(T^+S^-)^2=LT^+S^-,~~(S^-T^+)^2=LS^-T^+.
 \label{newproj}
 \end{equation}

 For the operators $S^{\pm (2)},~T^{\pm (2)}$ we first write
 \begin{eqnarray}
 S^{+(2)}&=&\sum_{p_1,p_2}A^{(l)}(p_1,p_2)
 \eta^{(l)\dagger}_{p_1}\eta^{(l)\dagger}_{p_2}\nonumber\\
 T^{+(2)}&=&\sum_{p_1,p_2}A^{(l)*}(-p_1,-p_2)
 \eta^{(l)\dagger}_{p_1}\eta^{(l)\dagger}_{p_2}\nonumber\\
 S^{-(2)}&=&-\sum_{p_1,p_2}A^{(l)}(-p_1,-p_2)
 \eta^{(l)}_{p_1}\eta^{(l)}_{p_2}\nonumber\\
 T^{-(2)}&=&-\sum_{p_1,p_2}A^{(l)*}(p_1,p_2)
 \eta^{(l)}_{p_1}\eta^{(l)}_{p_2}
 \label{steta}
 \end{eqnarray}
 where
 \begin{equation}
 A^{(l)}(p_1,p_2)
 ={ie^{-\pi i S^z}\over 2L}\sum_{1\leq j<k\leq L}\{e^{-ij({\pi\over
 2}+p_1)}e^{-ik({\pi\over 2}+p_2)}-
 e^{-ij({\pi\over
 2}+p_2)}e^{-ik({\pi\over 2}+p_1)}\}.
 \label{appdef}
 \end{equation}
 We have antisymmetrized this expression because of the anticommutation
 relations (\ref{etacomm}).
 However
 to proceed further we need to know if for some $p$ we can have
 \begin{equation}
 p+{\pi\over 2}\equiv 0 ({\rm mod}~2\pi).
 \label{pio2}
 \end{equation}
 The possibility of this holding depends both on $l$ and on whether or
 not $L/2$ is even or odd.

 Consider first the case where there is no possible value of $p$ for which 
 (\ref{pio2}) can hold. We see from (\ref{momentum1}) that this occurs
 either if $l=0$ and $L/2$ is even or $l=1$ and $L/2$ is odd. 
 We see from (\ref{hldef}) that in this case the operators
 $\eta^{(l)}_p$ are the operators in $H_l$ for $S^z\equiv 0 ({\rm mod}~2)$
 Then we do the sum over $j$ in (\ref{appdef}) and find  
 \begin{eqnarray}
 A^{(l)}={ie^{-\pi i S^z}\over 2L}\sum_{1\leq k \leq L}\{& & e^{-ik({\pi\over
 2}+p_2)}{e^{-i({\pi\over 2}+p_1)}-e^{-ik({\pi\over 2}+p_1)}\over
 1-e^{-i({\pi\over 2}+p_1)}}\nonumber\\
 &- & e^{-ik({\pi\over
 2}+p_1)}{e^{-i({\pi\over 2}+p_2)}-e^{-ik({\pi\over 2}+p_2)}\over
 1-e^{-i({\pi\over 2}+p_2)}}\}.
 \label{step1}
 \end{eqnarray}
 The sum on $k$ is now also a geometric series whose value depends on
 whether or not
 \begin{equation}
 \pi +p_1+p_2\equiv 0 ({\rm mod}~2\pi)
 \label{picond}
 \end{equation} 
 which is allowed in both cases $l=0$ with $L/2$ even and $l=1$ with
 $L/2$ odd even though (\ref{pio2}) can not hold. If there are no $p$
 which satisfy (\ref{picond}) it is easy to see that since
 $e^{iL({\pi\over 2}+p)}=-1$ holds in both cases that
 the sum over $j$
 vanishes. However if (\ref{picond}) does hold the sum does not vanish
 and hence we find
 \begin{equation}
 A^{(l)}(p_1,p_2)=-{e^{-i \pi S^z}\over 2}\cot{1\over 2}(p_1+{\pi
 \over2})\delta_{p_1+p_2+\pi,0}.
 \label{finala}
 \end{equation}
 Thus we explicitly find from (\ref{steta})
 \begin{eqnarray}
 S^{+(2)}&=&-{e^{-i\pi S^z}\over 2}
 \sum_{p}\cot{1\over 2}(p+{\pi\over 2})
 \eta^{(l)\dagger}_{p}
 \eta^{(l)\dagger}_{\pi-p}\nonumber\\
 T^{+(2)}&=&-{e^{i\pi S^z}\over 2}\sum_{p}\tan{1\over 2}(p+{\pi\over 2})
 \eta^{(l)\dagger}_{p}\eta^{(l)\dagger}_
 {\pi-p}\nonumber\\
 S^{-(2)}&=&{e^{-i\pi S^z}\over 2}\sum_{p}\tan{1\over 2}(p+{\pi\over 2})
 \eta^{(l)}_{p}\eta^{(l)}_{\pi-p}\nonumber\\
 T^{-(2)}&=&{e^{i\pi S^z}\over 2}\sum_{p}\cot{1\over 2}(p+{\pi\over 2})
 \eta^{(l)}_{p}\eta^{(l)}_{\pi-p}.
 \label{stetafinal}
 \end{eqnarray}
 where we note that $[\eta^{(l)}_p\eta^{(l)}_{\pi-p},e^{i\pi S^z}]=0$

 The degeneracy of the spectrum of H in the sector
 $S^z\equiv 0~(\rm{mod}~2)$ where $p\neq \pi/2,~3\pi/2$
  presented in sec. 2 is now very
 transparently demonstrated. The key to this demonstration  is to
 consider the operators 
 \begin{equation}
 \eta^{(l)}_p\eta^{(l)}_{\pi-p},~~~~
 \eta^{(l)\dagger}_p\eta^{(l)\dagger}_{\pi-p} 
 \end{equation}
 which appear in the summands of the representations of the operators
 $S^{\pm(2)},~T^{\pm(2)}.$
 Using the representation of the Hamiltonian
 (\ref{hldef}) and the commutation relations (\ref{etacomm}) we find
 \begin{eqnarray}
 [\eta^{(l)\dagger}_p \eta^{(l)\dagger}_{\pi-p},H_l]&=&
 \cos p [\eta^{(l)\dagger}_p\eta^{(l)\dagger}_{\pi-p},
 \eta^{(l)\dagger}_{p}\eta^{(l)}_p]+
 \cos(\pi-p) [\eta^{(l)\dagger}_p\eta^{(l)\dagger}_{\pi-p},
 \eta^{(l)\dagger}_{\pi-p}\eta^{(l)}_{\pi-p}]\nonumber \\
 &=&-\cos p \eta^{(l)\dagger}_p\eta^{(l)\dagger}_{\pi-p}-\cos(\pi-p)
 \eta_p^{(l)\dagger}\eta_{\pi-p}^{(l)\dagger}=0
 \label{zeroenergy}
 \end{eqnarray}
 Thus the pair operator $\eta^{(l)\dagger}_p\eta^{(l)\dagger}_{\pi-p}$
 adds an excitation of zero energy with $S^z=2$ (and momentum $\pi$) to the
 system. Furthermore
 \begin{equation}
 (\eta_p^{(l)\dagger}\eta^{(l)\dagger}_{\pi-p})^2=0
 \label{spinhalf}
 \end{equation}
 so that only one pair is allowed for each value of $p.$ The properties
 (\ref{zeroenergy}) and (\ref{spinhalf}) reproduce the degeneracy found
 in sec. 2.

 It is now a simple matter to use (\ref{etacomm}) and
 (\ref{stetafinal}) to compute the commutation relations needed for the
 loop algebra of $sl_2$ given in (\ref{one}),(\ref{serreone})-(\ref{two}).
 We first obtain the single commutators
 \begin{eqnarray}
 {[}S^{+(2)},T^{+(2)}{]}&=&{[}S^{-(2)},T^{-(2)}{]}=0,\label{ocomm1}\\
 {[}S^{+(2)},S^{-(2)}{]} &=& [T^{-(2)},T^{-(2)}] 
 = \sum_{p}  (\eta^{(l)\dagger}_{p}\eta^{(l)}_{p} - \frac{1}{2})=S^z,\label{ocomm2}\\
 {[}S^{+(2)},T^{-(2)}{]} &=& \sum_{p} \cot^2{1\over 2}(p+{\pi\over  2})
  (\eta^{(l)\dagger}_{p}\eta^{(l)}_{p} -\frac{1}{2}),\label{ocomm3}\\
 {[}T^{+(2)},S^{-(2)}{]}& =& \sum_{p} \tan^2{1\over 2}(p+{\pi\over 2})
  (\eta^{(l)\dagger}_{p}\eta^{(l)}_{p} -\frac{1}{2}),
 \end{eqnarray}
 then the double commutators
 \begin{eqnarray}
 {[}T^{-(2)},[T^{-(2)},S^{+(2)}]{]}& =& -e^{\pi i S^z}\sum_{p} 
 \cot^3{1\over 2}(p+{\pi\over 2})
  \eta^{(l)}_{p}\eta^{(l)}_{\pi-p},\label{dcomm1}\\
 {[}T^{+(2)},[T^{+(2)},S^{-(2)}]{]}& =& e^{\pi i S^z}\sum_{p} 
 \tan^3{1\over 2}(p+{\pi\over 2})
  \eta^{(l)\dagger}_{p}\eta^{(l)\dagger}_{\pi-p},\label{dcomm2}\\
 {[}T^{+(2)},[T^{-(2)},S^{+(2)}]{]} &=& 2S^{+(2)}\\
 {[}T^{-(2)},[T^{+(2)},S^{-(2)}]{]}& =& 
 2S^{-(2)},
 \end{eqnarray}
 and finally the triple commutators
 \begin{eqnarray}
 {[}T^{-(2)},[T^{-(2)},[T^{-(2)},S^{+(2)}]]{]}&=&
  {[}T^{+(2)},[T^{+(2)},[T^{+(2)},S^{-(2)}]]{]}=0,\label{tcomm1}\\
 {[}S^{-(2)},[S^{-(2)},[S^{-(2)},T^{+(2)}]]{]}&=&
  {[}S^{+(2)},[S^{+(2)},[S^{+(2)},T^{-(2)}]]{]}=0,\label{tcomm2}\\
 {[}T^{+(2)},[T^{-(2)},[T^{-(2)},S^{+(2)}]]{]}& =&
  {[}T^{-(2)},[T^{+(2)},[T^{-(2)},S^{+(2)}]]{]} =2[T^{-(2)},S^{+(2)}]
  \label{tcomm3}\\
  {[}T^{+(2)},[T^{-(2)},[T^{+(2)},S^{-(2)}]]{]}& =&
  {[}T^{-(2)},[T^{+(2)},[T^{+(2)},S^{-(2)}]]{]}
  =2[T^{+(2)},S^{-(2)}]\label{tcomm4}
 \end{eqnarray}
 where the single commutators(\ref{ocomm1}) and (\ref{ocomm2}) are
 (\ref{one}) and (\ref{two}). 
 The triple commutators (\ref{tcomm1}),(\ref{tcomm2}) are the
 Serre relations (\ref{serreone})-(\ref{serrefour}) of the algebra and
 follow immediately from (\ref{stetafinal}), (\ref{dcomm1}) and
 (\ref{dcomm2}) by use of (\ref{spinhalf}).

 On the other hand if $l=0$ with $L/2$ odd or $l=1$ with $L/2$ even then
 the condition (\ref{pio2}) can hold. In both cases we see from
 (\ref{hldef}) that the operators $\eta^{(l)}_p$ are those of $H_l$
 with $S^z\equiv 1~({\rm mod}~2).$ There are now two cases to
 consider depending on whether or not (\ref{pio2}) is satisfied.

 If there are no values of $p$ which satisfy (\ref{pio2}) then by using
 $e^{iL({\pi\over 2}+p)}=1$ we find as before that $A^{(l)}(p_1,p_2)$
 is given by (\ref{finala}). But if either $p_1$ or $p_2$ is $3\pi/2$
 and $p_1+p_2+\pi\neq 0~({\rm mod} 2\pi)$ we find that
 $A^{(l)}(p_1,p_2)$ does not vanish and that instead we have
 \begin{equation}
 A^{(l)}({3\pi\over 2},p)=-A^{(l)}(p,{3\pi\over 2})=-{ie^{-\pi i
 S^z}\over 1-e^{-i({\pi \over 2}+p)}}.
 \end{equation}
 Thus instead of (\ref{stetafinal}) we have
 \begin{eqnarray}
 S^{+(2)}&=&{e^{-iS^z}\over 2}\sum_{p\neq 3\pi/2}(\eta^{(l)\dagger}_{p}
 \eta^{(l)\dagger}_{\pi-p}\cot{1\over 2}(p+{\pi\over 2})
 -2\eta_{3\pi/2}^{(l)\dagger}\eta_p^{(l)\dagger}{i\over
 1-e^{-i({\pi\over 2}+p)}})
 \label{newfinal}
 \end{eqnarray}

 The second term in (\ref{newfinal})      does not commute with the
 Hamiltonian and as discussed in sec. 2 we need instead
 to use the projection operators (\ref{newproj}) to construct the
 projected operators 
 \begin{eqnarray}
 S^{+(2)}_{pr}=S^{+(2)}S^+T^-+T^-S^+S^{+(2)},\nonumber\\
 S^{-(2)}_{pr}=S^{-(2)}S^-T^++T^+S^-S^{-(2)},\nonumber\\
 T^{+(2)}_{pr}=T^{+(2)}T^+S^-+S^-T^+T^{+(2)},\nonumber\\
 T^{-(2)}_{pr}=T^{+(2)}T^-S^++S^+T^-T^{-(2)}.
 \end{eqnarray}
 These projected operators are readily expressed in terms of
 $\eta_p^{(l)}$ by use of (\ref{newfinal}) and the identity 
 \begin{equation}
 \eta_p\eta_p^{\dagger}\eta_{p_1}^{\dagger}\eta_{p_2}^{\dagger}+
 \eta^{\dagger}_{p1}\eta_{p_2}^{\dagger}\eta_{p}^{\dagger}\eta_{p}=
 \eta_{p1}^{\dagger}\eta_{p_2}^{\dagger}(1-\delta_{p,p_1}-\delta_{p,p_2})
 \end{equation}
 Thus we find that the noncommuting terms of
 $\eta_{3\pi/2}^{\dagger}\eta_p$ and $\eta_{\pi/2}^{\dagger}\eta_p$ are
 annihilated by the projection operator and we obtain the expressions
 analogous to the unprojected expressions (\ref{stetafinal})
 \begin{eqnarray}
 S_{pr}^{+(2)}&=&-{e^{-i\pi S^z}\over 2}\sum_{p\neq \pi/2,~3\pi/2}
 \cot{1\over 2}(p+{\pi\over2})
 \eta^{(l)\dagger}_{p}
 \eta^{(l)\dagger}_{\pi-p}\nonumber\\
 T_{pr}^{+(2)}&=&-{e^{i\pi S^z}\over 2}\sum_{p\neq \pi/2,~3\pi/ 2}
 \tan{1\over 2}(p+{\pi\over 2})
 \eta^{(l)\dagger}_{p}\eta^{(l)\dagger}_
 {\pi-p}\nonumber\\
 S_{pr}^{-(2)}&=&{e^{-i\pi S^z}\over 2}\sum_{p\neq \pi/2, 3 \pi /2}
 \tan{1\over 2}(p+{\pi\over 2})
 \eta^{(l)}_{p}\eta^{(l)}_{\pi-p}\nonumber\\
 T_{pr}^{-(2)}&=&{e^{i\pi S^z}\over 2}
 \sum_{p\neq \pi/2, 3\pi/2}
 \cot{1\over 2}(p+{\pi\over 2}).
 \eta^{(l)}_{p}\eta^{(l)}_{\pi-p}
 \label{prstetafinal}
 \end{eqnarray}
 These operators manifestly commute with the Hamiltonian and the
 computation of the degeneracy is identical with the previous case of
 $S^z\equiv 0~({\rm mod}~2).$ Similarly all the commutation relations
 of the $sl_2$ loop algebra (\ref{ocomm1})-(\ref{tcomm4}) 
 hold for the projected operators with
 the one exception of (\ref{ocomm2}) where we find the slight modification
 \begin{equation}
 {[}S^{+(2)},S^{-(2)}{]} = [T^{-(2)},T^{-(2)}] 
 = \sum_{p \neq \pi/2, 3\pi/2}  (\eta^{(l)\dagger}_{p}\eta^{(l)}_{p} - \frac{1}{2})
 =S^z-(\eta^{(l)\dagger}_{\pi/2}\eta^{(l)}_{\pi/2} - \frac{1}{2})-
 (\eta^{(l)\dagger}_{3 \pi/2}\eta^{(l)}_{3 \pi/2} - \frac{1}{2})
 ,\label{procomm2}
 \end{equation}

 \section{Discussion}

 In sec. 4 we have demonstrated that the operators $S^{\pm(2)}$ and
 $T^{\pm(2)}$ obtained from the quantum group $U_q(sl_2)$ at $q=e^{i\pi/2}$
 both explains the degeneracies of the spectrum of the Hamiltonian
 found in sec. 2 and obey the defining commutation relations of the loop
 algebra $sl_2$ given in sec. 3. The reason for this more elaborate
 treatment of the degeneracies computed by elementary means in sec. 2 is
 that for $N>2$ where the treatment of sec. 2 no longer is possible we
 have found that the $sl_2$ symmetry algebra of sec. 2 still holds and
 therefore the space of eigenstates of the Hamiltonian (\ref{hxxz})
 decomposes into a direct sum of evaluation representations of the loop
 algebra of $sl_2.$ This decomposition is explicitly contained in the
 representation given in (\ref{stetafinal}) for the unprojected
 operators and (\ref{prstetafinal}) for their projected counterparts.
 In this expression the fact that 
 $(\eta_p^{\dagger}\eta_{\pi-p}^{\dagger})^2=0$
 is equivalent to the statement that only spin 1/2 representations
 occur.

 The demonstration that only spin 1/2 representations occur is
 more complicated than the demonstration of the $sl_2$ loop algebra
 symmetry. For $N=2$ we can in principle compute sufficiently
 many multiple commutators and show that they are of the form of the
 right hand side of (\ref{dcomm2})
 with $\tan ^3(p+\pi/2)/2$ replaced by various other powers. Thus we
 can generate $L$ equations for the $L$ different operators
 $\eta^{\dagger}_p\eta^{\dagger}_{\pi-p}$ and solve the system.
 In practice what we did in section 4 was to examine
 the summands which appeared in the expressions for $S^{\pm(2)}$ and
 $T^{\pm(2)}$and then discovered that these summands also commuted with
 the Hamiltonian. The ability to do this for $N\geq 3$
 relies on having a proper form for $S^{\pm (N)}.$

 In this paper the proper form relied on the Jordan--Wigner operators
 and thus it seems profitable to generalize the fermionic   operators $c_j$
 to ``parafermionic'' operators in position space 
 \begin{eqnarray}
 c_j&=&q^{2\sum_{k=1}^{j-1}\sigma_k^{+}\sigma_k^-}\sigma^-_j
 =q^{j-1}q^{\sum_{k=1}^{j-1}\sigma^z_k}\sigma^-_j\nonumber\\
 c_j^*&=&q^{-2\sum_{k=1}^{j-1}\sigma_k^{+}\sigma_k^-}\sigma^-_j
 =q^{-(j-1)}q^{-\sum_{k=1}^{j-1}\sigma^z_k}\sigma^-_j\nonumber\\
 c^{\dagger}_j&=&q^{-2\sum_{k=1}^{j-1}\sigma_k^{+}\sigma_k^-}\sigma^+_j
 =q^{-(j-1)}q^{-\sum_{k=1}^{j-1}\sigma^z_k}\sigma^+_j\nonumber\\
 c^{*\dagger}_j&=&q^{2\sum_{k=1}^{j-1}\sigma_k^{+}\sigma_k^-}\sigma^+_j
 =q^{j-1}q^{\sum_{k=1}^{j-1}\sigma^z_k}\sigma^+_j
 \end{eqnarray}
 This generalization can be carried out along the lines of the
  treatment of parafermions in ref. \cite{zamfat}
  but gives somewhat cumbersome
 expressions for the (anti) commutators of the generalization of the
 $\eta_p.$

 A second possibility is to consider instead of the Jordan-Wigner
 operators the Bethe's ansatz wave function as given (for example) by
 Yang and Yang \cite{yy}. This wave function is of the form
 \begin{equation}
 \psi=\sum_P A_P e^{i\sum_j k_{Pj}x_j}
 \end{equation}
 where the sum is over all permutations $P$ and
 \begin{equation}
 A_P(2\Delta e^{ip_1}-1-e^{i(p_1+p_2)})=A_{P'}(2\Delta e^{ip_2}-1-e^{i(p_1+p_2)})
 \label{condition}
 \end{equation}
 for permutations $P$ and $P'$ which differ only in the interchange of
 the two adjacent elements
 $p_1$ and $p_2$. If we naively set $\Delta=0$ we find $A_P=A_{P'}$ and
 thus $\psi$ is a Slater determinant  wave function.

 Now the coordinate space form of the Jordan Wigner wave function is
 also a Slater determinant and thus it might be expected that the
 Bethe's wavefunctions at $\Delta=0$ and the Jordan--Wigner
 wavefunction are identical. It is most important to realize, however,
 that this is in fact not the  case when we are considering degenerate
 eigenvalues. This can be seen very explicitly by considering the
 operation of the spin reflection operator $R$ on the states with $S^z=0$.
 All the Bethe's wave functions are eigenstates of $R$ for $\Delta \neq
 0$ and thus continue to be eigenfunctions at $\Delta=0.$ But a direct
 computation shows that the Jordan Wigner states are not eigenstates
 of $R$. Moreover if the solutions of the Bethe's equations of ref.
 \cite{yy} for $\Delta\neq 0$ are smoothly continued to $\Delta=0$ we
 have explicitly found for the degenerate eigenvalues that there
 are complex solutions to Bethe's equations which remain complex even in
 the limit $\Delta=0$.

 Part of what is happening is that in the degenerate subspace there are
 solutions with pairs $p_1$ and $p_2$ where $p_1+p_2=\pi.$ But when this
 condition is put into (\ref{condition}) and then we set $\Delta=0$ we
 see that (\ref{condition}) reduces to $0=0$ and  the relation
 between $A_P$ and $A_{P'}$ is no longer determined. Indeed it is
 exactly this loophole in the argument of ref.\cite{yy} which is
 exploited by Baxter \cite{baxc}-\cite{baxe}in his 
 computation of the eigenvectors of the XYZ
 model at roots of unity. 

 We thus conclude that for degenerate eigenstates the Jordan Wigner
 states are linear combinations of the Bethe's states and that  the
 mechanism needed to recognize the spin 1/2 nature of the
 representations will depend on which set of basis states is used.

We note further that a numerical study of the eigenvalues of the XYZ
spin chain indicates that when the root of unity condition $q^{2N}=1$
is generalized to Baxter's condition ((C15) of ref.\cite{baxb}) of
\begin{equation}
2N\eta=2m_1K+im_2K'
\end{equation}
that the size of the degenerate multiplets of the XXZ model are 
diminished by at most a factor of two. This clearly indicates that
with an appropriate generalization of the operators $S^{\pm(N)}$ and
$T^{\pm(N)}$ the $sl_2$ loop symmetry of the XXZ model extends
to the XYZ model.

 \bigskip

 {\bf Acknowledgments}

 \bigskip

 We wish to  thank M. Batchelor,
 R.J. Baxter, V.V. Bazhanov, E. Date, B. Davies, A. Kirillov, P.A. Pearce,
 J.H.H.Perk and S-S Roan for useful discussions. We wish to particularly
 thank M. Jimbo for calling our attention to the results of reference
 (\cite{lusc}) on higher order Serre relations which allows us to
 markedly simplify our proof of the Serre relations
 (\ref{serreone})-(\ref{serrefour}).
 One of us (BMM)   
 is pleased to thank M. Batchelor for hospitality extended to the
 Australian National University where part of this work was done. 
 This work is supported in part by the National Science Foundation
 under Grant No. DMR--9703543

 \bigskip 
 \appendix


\section{Commutation (anti-commutation) relations with the transfer matrix } 

In this appendix  we  show that 
the transfer matrix $T(v)$ 
of the six-vertex model can be written 
as a sum of products of the Temperley-Lieb generators 
multiplied by the shift operators and furthermore that for all $q$ 
the operators $S^{\pm}$ and $T^{\pm}$ commute with the Temperley--Lieb
operators.
Thus, any operator constructed from  $S^{\pm}$ and $T^{\pm}$ 
commutes (anti-commutes) with the transfer matrix 
if it (anti-)commutes with the shift operators. Therefore
since we proved in section 3.E that $S^{\pm (N)}$ and $T^{\pm (N)}$
(anti)-commutes with the shift operator 
and that the operators given by (\ref{proj}) 
and (\ref{Nproj}) (anti)-commute with the shift operator   
the (anti-)commutation with the full transfer
matrix follows.   

\subsection{Twisted transfer matrix of the six-vertex model}

\par 
There are many different sets of local Boltzmann weights which give
the same transfer matrix. These Boltzmann weights differ by a gauge
transformation. In this appendix we find it convenient to use two
different gauge equivalent sets of Boltzmann weights 
\cite{Akutsu-Wadati} which we denote by
$W^+(\mu,\nu )|_{\alpha,\beta}$ and  $W^-(\mu,\nu )|_{\alpha,\beta}$
whose the nonzero elements are given   by
 \begin{eqnarray}
 { W^\pm}(1,1)|_{1,1}&=&{ W^\pm}(-1,-1)|_{-1,-1}=
 - 2 \sinh v
  \label{weight1} \\
 {W^\pm}(-1,-1)|_{1,1}&=&{W^\pm}(1,1)|_{-1,-1}
 = 2 \sinh(\lambda-v)\label{weight2} \\  
 {W^\pm}(-1,1)|_{1,-1}&=& 2 e^{\pm v} \sinh \lambda  \label{weight3} \\
 {W^\pm}(1,-1)|_{-1,1}& =& 2 e^{\mp v} \sinh \lambda 
 \label{weight4} 
 \end{eqnarray}
where $q=e^{\lambda}$ and we recall that $\Delta=(q+q^{-1})/2$.   
More generally we will also need the ``twisted'' Boltzmann weights 
 with the twisting parameter $\phi$ by 
\begin{equation} 
{\tilde W}^{\pm}(\mu,\nu;\phi)|_{\alpha,\beta} = 
q^{\phi(\mu+\alpha)} \, W^{\pm}(\mu,\nu)|_{\alpha,\beta}  
, \quad {\rm for} \quad 
\mu,\nu, \alpha, \beta = \pm 1\, .   
\end{equation} 
The matrix elements 
of  the ``twisted'' transfer matrix $T(v;\phi)$ are then defined by  
\begin{eqnarray} 
& & \left(T(v; \phi)\right)^{\mu_1, \cdots, \mu_L}_{\nu_1, \cdots, \nu_L}  
 = {\rm Tr} \, {\tilde W}^\pm (\mu_1, \nu_1;\phi) W^\pm (\mu_2, \nu_2) 
\cdots W^\pm (\mu_L, \nu_L) \nonumber \\ 
 && = \sum_{\alpha_1 \cdots \alpha_L} 
{\tilde W}^\pm(\mu_1, \nu_1;\phi)|_{\alpha_1, \alpha_2}  
W^\pm(\mu_2, \nu_2)|_{\alpha_2, \alpha_3}  
\cdots 
W^\pm(\mu_L, \nu_L)|_{\alpha_L, \alpha_1} 
\label{twistT}
\end{eqnarray} 
 Due to the "{\it ice rule}", we have  the same number of configurations 
 for the two weights $ W^{\pm}(-1,1)|_{1,-1}$ and  
 $W^\pm (1,-1)|_{-1,1}$ in the product (\ref{twistT})  
 and thus the changes in the Boltzmann weights are cancelled  out for 
 the twisted transfer matrix $T(v;\phi)$. 

We denote the twisted transfer matrix at $\phi=0$ by
$T(v)=T(v; \phi=0)$ and note that the transfer matrix $T(v)$ 
is related to the shift operator 
$\Pi_L$ (\ref{shift}) and the XXZ Hamiltonian (\ref{hxxz}) by
\begin{equation} 
T(v \approx 0)  =  (q-q^{-1})^L \, \Pi_L \,  
\left\{ I   - {2v \over  {q- q^{-1}}} 
\left( H +  {{L(q+q^{-1})} \over 4} \right)   + o (v) \right\} 
\end{equation} 
We note in passing that just as the sign of the Hamiltonian
(\ref{hxxz}) may be negated by a similarity transformation that same
transformation changes the sign of the Boltzmann weight
(\ref{weight1})  
sends $T(v) \rightarrow (-1)^{L/2-S^z} U T(v) U^{-1}$ 
where $U= \sigma_1^z \otimes I_2 \otimes \cdots \otimes \sigma_{L-1}^z 
\otimes I_L$.

\subsection{Operators $X^{\pm}_j(v)$ and 
a decomposition of the transfer matrix}

\par 
We define ${\tilde X}^{\pm}_j(v;\phi)$ for $j=1, \ldots, L-1$,    
by the following 
\begin{equation} 
{\tilde X}^\pm_j(v;\phi) = 
\sum_{a,b,c,d=\pm 1 } {\tilde X}^{\pm ac}_{~~ bd}(v;\phi) 
I_1 \otimes \cdots \otimes I_{j-1} \otimes E^{ab}_{j} \otimes E^{cd}_{j+1}  
\otimes I_{j+2} \otimes \cdots \otimes I_L  
\label{tensor} 
\end{equation} 
and define
\begin{equation}
X^{\pm}_j(v)={\tilde X}^{\pm}_j(v;0)
\end{equation}
where $E^{ab}$ denotes the matrix  
\begin{equation} 
\left( E^{ab} \right)_{c, d} 
= \delta_{a,c } \, \delta_{b, d} \quad {\rm for} \quad  c , d = \pm 1 \, .  
\end{equation}
and
\begin{eqnarray} 
{\tilde X}^{\pm ac}_{~~bd}(v;\phi) 
& = & q^{\phi(a-b)} 
\, X^{\pm ac}_{~~bd}(v)  
=  q^{\phi(a-b)}  W^{\pm} (a, d)|_{{-b},{-c}} 
\nonumber \\ 
& =&  2\sinh(\lambda-v) \delta_{a,b} \delta_{c,d} 
+ 2 \sinh v ~ a b e^{\mp(a+b)\lambda/2} \, 
q^{\phi(a-b)} \, 
\delta_{a, {-c}} 
\delta_{b, {-d}} \,  
\label{Xabcd_twist} 
\end{eqnarray} 
where we note the symmetry \cite{Deguchi} 
 \begin{equation}
 {X}^{\pm \, a,c}_{~~\, b,d}(v) 
 = -ade^{\mp(a-d)\lambda/2}{X}^{\pm \, c, -{d}}_{~~\, -{a}, b }(\lambda-v)  
 = -bce^{\mp(b-c)\lambda/2} {X}^{\pm \, -{b}, a}_{~~\, d, -{c} }(\lambda-v)  
 \label{crossing} 
 \end{equation}
Thus we find that the expression (\ref{twistT}) for the transfer matrix
becomes
 \begin{eqnarray} 
 \left( T(v; \phi) \right)^{\mu_1 \mu_2 \ldots \mu_L} 
 _{\nu_1 \nu_2 \ldots \nu_L} 
 & = & \sum_{\alpha_1, \ldots, \alpha_L} 
 {\tilde W}^{\pm}
(\mu_1,\nu_1;\phi)|_{\alpha_1, \alpha_2} 
 W^{\pm}(\mu_2,\nu_2)|_{\alpha_2, \alpha_3} 
 \cdots  W^{\pm}(\mu_L,\nu_L)|_{\alpha_L, \alpha_1} 
 \nonumber \\ 
 & = & \sum_{\alpha_1, \ldots, \alpha_L} 
 {\tilde X}^{\pm \mu_1, -{\alpha}_2}_{~~ -{\alpha}_1, \nu_1}(v;\phi)  
 X^{\pm \mu_2, -{\alpha}_3}_{~~ -{\alpha}_2, \nu_2}(v)  
 \cdots  X^{\pm \mu_L,-{\alpha}_1}_{~~ -{\alpha}_L, \nu_L}(v) 
 \nonumber \\ 
 & = & \sum_{\beta_0, \ldots, \beta_{L-1}} 
 X^{\pm \mu_L,{\beta}_0}_{~~ {\beta}_{L-1}, \nu_L}(v) 
 \cdots  
 X^{\pm \mu_2, {\beta}_2}_{~~{\beta}_1, \nu_2}(v)  
 {\tilde X}^{\pm \mu_1, {\beta}_1}_{~~{\beta}_0, \nu_1}(v;\phi)  
 \nonumber \\ 
 &=& \sum_{\beta_0, \beta_1} 
\left( 
 X^{\pm}_{L-1}(v) \cdots X^{\pm}_1(v)   
 \right)^{\mu_2 \mu_3 \cdots \mu_L \beta_0}
 _{\beta_1 \nu_2 \cdots \nu_{L-1} \nu_L} \,   
 {\tilde X}^{\pm \mu_1 \beta_1}_{~~ \beta_0 \nu_1}(v;\phi) 
 \label{TXXX}
 \end{eqnarray} 

 \par 
 Substituting (\ref{Xabcd_twist})
  into the first factor ${\tilde X}^{\pm \mu_1 \beta_1}_{~~\beta_0 \nu_1}$ 
 of  the last line of (\ref{TXXX}), we have 
 \begin{eqnarray} 
 \left( T(v; \phi) \right)^{\mu_1 \cdots \mu_L}_{\nu_1 \cdots \nu_L}  
 %
 %
& = &  2 \sinh(\lambda-v) 
 \sum_{\beta_0, \beta_1}
 \delta_{\mu_1,\beta_0} \delta_{\beta_1,\nu_1}   
 \left( 
 X^{\pm}_{L-1}(v) X^{\pm}_{L-2}(v) \cdots X^{\pm}_{1}(v)  
 \right)^{\mu_2 \cdots \mu_L \beta_0}_{\beta_1 \nu_2 \cdots \nu_L}  
 \nonumber \\ 
& &  +  2 \sinh v \sum_{\beta_0, \beta_1}
 \mu_1 \beta_0 e^{\mp(\mu_1+\beta_0)\lambda/2} 
q^{\phi (\mu_1-\beta_0)} \, 
 \delta_{\mu_1,-\beta_1} 
 \delta_{\beta_0, -\nu_1} \,  
\nonumber \\
& & \qquad \times 
 \left( 
 X^{\pm}_{L-1}(v) X^{\pm}_{L-2}(v) \cdots X^{\pm}_{1}(v)  
 \right)^{\mu_2 \cdots \mu_L \beta_0}_{\beta_1 \nu_2 \cdots \nu_L}  
 \nonumber \\ 
& = &  2 \sinh(\lambda-v) \, 
 \left( 
 X^{\pm}_{L-1}(v) X^{\pm}_{L-2}(v) \cdots X^{\pm}_{1}(v)  
 \right)^{\mu_2 \cdots \mu_L \mu_1}_{\nu_1 \nu_2 \cdots \nu_L}  
 \nonumber \\ 
& &  -  2 \sinh v \, 
 \mu_1\nu_1e^{\mp(\mu_1-\nu_1)\lambda/2}
%
%
 q^{\phi (\mu_1+\nu_1)}
 \left( 
 X^{\pm}_{L-1}(v) X^{\pm}_{L-2}(v) \cdots X^{\pm}_{1}(v)  
 \right)^{\mu_2 \cdots \mu_L, -{\nu}_1}_{-{\mu}_1 \nu_2 \cdots \nu_L}  
 \nonumber \\ 
& = &  2 \sinh(\lambda-v) \, 
 \left( 
 \Pi_L  X^{\pm}_{L-1}(v) X^{\pm}_{L-2}(v) \cdots X^{\pm}_{1}(v)  
 \right)^{\mu_1 \mu_2 \cdots \mu_L}_{\nu_1 \nu_2 \cdots \nu_L}  
 \nonumber \\ 
& &  +  2 \sinh v
 \left( q^{\phi \sigma_1^z} \, 
 X^{\pm}_{1}(\lambda-v) X^{\pm}_{2}(\lambda-v) \cdots 
X^{\pm}_{L-1}(\lambda-v)  
 \Pi_R \, q^{\phi \sigma_1^z} \right)^{\mu_1 \cdots \mu_{L-1} \mu_L}_
{\nu_1 \cdots \nu_{L-1} \nu_L}  
 \end{eqnarray} 
 Here  we have made use of the following relations    
 \begin{eqnarray} 
& &  
 \left( 
 X^{\pm}_{L-1}(v) X^{\pm}_{L-2}(v) \cdots X^{\pm}_{1}(v)  
 \right)^{\mu_2 \cdots \mu_L \mu_1}_{\nu_1 \nu_2 \cdots \nu_L}  
\nonumber \\
& & \quad =\left( 
  \Pi_L 
  X^{\pm}_{L-1}(v) X^{\pm}_{L-2}(v) \cdots X^{\pm}_{1}(v)  
 \right)^{\mu_1 \mu_2 \cdots \mu_L}_{\nu_1 \nu_2 \cdots \nu_L}  
 \label{term1}
 \end{eqnarray}
 and 
 \begin{eqnarray} 
  & & 
 -\mu_1\nu_1e^{\mp(\mu_1-\nu_1)\lambda/2} q^{\phi (\mu_1+\nu_1)} 
 \left( 
 X^{\pm}_{L-1}(v) X^{\pm}_{L-2}(v) \cdots X^{\pm}_{1}(v)  
 \right)^{\mu_2 \cdots \mu_L, -{\nu}_1}_{-{\mu}_1, \nu_2 \cdots \nu_L}  
 \nonumber \\ 
 & & \quad =  
 \left( q^{\phi \sigma_1^z} \,  
 X^{\pm}_{1}(\lambda-v) X^{\pm}_{2}(\lambda-v) \cdots 
 X^{\pm}_{L-1}(\lambda-v) 
  \Pi_R \, q^{\phi \sigma_1^z} 
 \right)^{\mu_1 \mu_2 \cdots \mu_{L}}_{\nu_1 \nu_2 \cdots \nu_L}  
 \label{term2} 
 \end{eqnarray} 
 The relation (\ref{term1}) is readily derived from the definition 
 (\ref{shift}) of the shift operator. 
 We can show 
 the relation (\ref{term2}) by making use of (\ref{crossing}) as follows.  
 \begin{eqnarray}
 & & 
 \left( 
 {X}^{\pm}_{L-1}(v) {X}^{\pm}_{L-2}(v) \cdots {X}^{\pm}_{1}(v)  
 \right)^{ \mu_2 \cdots \mu_L, -{\nu}_1}_{-{\mu}_1, \nu_2 \cdots \nu_L}  
 \, 
 (-\mu_1 \nu_1 e^{\mp(\mu_1-\nu_1)\lambda/2}) \, q^{\phi (\mu_1+\nu_1)} 
 \nonumber \\ 
 & &  = \sum_{\alpha_2, \cdots, \alpha_{L-1}} \, 
 (-\nu_1 e^{\pm \nu_1 \lambda/2}) 
 \, 
 {X}^{\pm \mu_L, -{\nu}_1}_{~~ \alpha_{L-1}, \nu_L}(v) 
 X^{\pm \mu_{L-1}, \alpha_{L-1}}_{~~ \alpha_{L-2}, \nu_{L-1}}(v)
 \cdots X^{\pm \mu_2, \alpha_2}_{~~ -{\mu}_1, \nu_2}(v) \,  
 (\mu_1 e^{\mp \mu_1 \lambda/2}) \, q^{\phi (\mu_1+\nu_1)} 
 \nonumber \\ 
 & & =\sum_{\alpha_2, \cdots, \alpha_{L-1}} 
q^{\phi (\mu_1+\nu_1)}  (-\nu_1 e^{ \pm \nu_1 \lambda/2})  
 \cdot \, 
(\alpha_{L-1} \nu_1 e^{\mp(\alpha_{L-1} + \nu_1)\lambda/2})
 X^{\pm \, -{\alpha}_{L-1}, \mu_L}_{~~\, \nu_L, \nu_1}(\lambda-v) \cdot  
\nonumber \\
& & \qquad \times 
(- \alpha_{L-2} \alpha_{L-1} e^{\mp(\alpha_{L-2} -\alpha_{L-1})\lambda/2} ) 
 X^{\pm \, -{\alpha}_{L-2}, \mu_{L-1}}_{~~ \, \nu_{L-1}, -{\alpha}_{L-1}}
(\lambda-v) 
 \cdots 
 \nonumber \\
 & & \qquad  
 \cdots 
 (-\alpha_2 \alpha_3 e^{\mp (\alpha_2 - \alpha_3)\lambda/2})
 X^{\pm -{\alpha}_2, \mu_3}_{~~ \nu_3, -{\alpha}_3} (\lambda-v) \cdot  
 ( \mu_1 \alpha_2 e^{\mp(-\mu_1 -\alpha_2)\lambda/2}) 
 X^{\pm \mu_1, \mu_2}_{~~ \nu_2, -{\alpha}_2} (\lambda-v) \cdot  
 \, 
 (\mu_1 e^{\mp \mu_1 \lambda/2})
 \nonumber \\ 
 & & = \sum_{\alpha_2, \ldots, \alpha_{L-1}} 
q^{\phi (\mu_1+\nu_1)} \cdot (-1)^L \cdot 
 X^{\pm -{\alpha}_{L-1}, \mu_L}_{~~ \nu_L {\nu}_1}(\lambda-v)   
 X^{\pm -{\alpha}_{L-2}, \mu_{L-1}}_{~~ \nu_{L-1}, -{\alpha}_{L-1}}(\lambda-v) 
 \cdots \nonumber \\ 
& & \qquad \qquad \times 
 X^{\pm -{\alpha}_2, \mu_3}_{~~ \nu_3, -{\alpha}_3} (\lambda-v) 
 X^{\pm \mu_1, \mu_2}_{~~ \nu_2, -{\alpha}_2} (\lambda-v)   
 \nonumber \\ 
 & & =
 \sum_{{\alpha}_2, \cdots, {\alpha}_{L-1}} 
q^{\phi (\mu_1+\nu_1)} \, 
 X^{\pm \mu_1, \mu_2}_{~~ \nu_2, -{\alpha}_2} (\lambda-v)   
 X^{\pm -{\alpha}_2, \mu_3}_{~~ \nu_3, -{\alpha}_3} (\lambda-v) 
\cdots 
 X^{\pm -{\alpha}_{L-2}, \mu_{L-1}}_{~~\nu_{L-1}, -{\alpha}_{L-1}}(\lambda-v) 
 X^{\pm -{\alpha}_{L-1}, \mu_L}_{~~ \nu_L, \nu_1}(\lambda-v) 
 \nonumber \\ 
 & & =
 \sum_{\beta_2, \cdots, \beta_{L-1}} 
q^{\phi (\mu_1+\nu_1)} \, 
 X^{\pm \mu_1, \mu_2}_{~~ \nu_2, \beta_2} (\lambda-v)  
 X^{\pm \beta_2, \mu_3}_{~~ \nu_3, \beta_3} (\lambda-v)  
 \cdots 
 X^{\pm \beta_{L-2}, \mu_{L-1}}_{~~ \nu_{L-1}, \beta_{L-1}}(\lambda-v) 
 X^{\pm \beta_{L-1}, \mu_L}_{~~ \nu_L, \nu_1}(\lambda-v) 
 \nonumber \\ 
 & & = 
q^{\phi (\mu_1+\nu_1)} \, 
 \left( 
 X^{\pm}_1(\lambda-v) 
 X^{\pm}_2(\lambda-v) \cdots  
 X^{\pm}_{L-1}(\lambda-v) 
 \right)^{\mu_1 \cdots \mu_{L-1} \mu_L}_{\nu_2 \cdots \nu_L \nu_1} 
 \nonumber \\
 & & =  \left( q^{\phi \sigma_1^z}\, 
 X^{\pm}_1(\lambda-v) 
 X^{\pm}_2(\lambda-v) \cdots  
 X^{\pm}_{L-1}(\lambda-v) 
 \Pi_R \, q^{\phi \sigma_1^z} \right)^{\mu_1 \cdots \mu_L}
_{\nu_1 \cdots \nu_L} 
 \end{eqnarray} 
where we used the fact that $L$ is even.

 \par 
 In summary, we have  
 \begin{equation}
 T(v; \phi)  =\Pi_L X_{LL}^{\pm}+X_{RR}^{\pm}\Pi_R  
 \label{twoterms}
\end{equation}
with 
\begin{eqnarray} 
X_{LL}^{\pm}&=&
2 \sinh(\lambda-v) \, 
  {X}^{\pm}_{L-1}(v) {X}^{\pm}_{L-2}(v) 
 \cdots {X}^{\pm}_{1}(v) 
 \nonumber \\ 
X_{RR}^{\pm}&=&
2 \sinh v  \, q^{\phi \sigma_1^z} \,  
   {X}^{\pm}_{1}(\lambda-v) {X}^{\pm}_{2}(\lambda-v) 
 \cdots {X}^{\pm}_{L-1}(\lambda-v)  \, q^{\phi \sigma_1^z} 
 \end{eqnarray}

 \subsection{The Temperley-Lieb algebra and the operators $S^{\pm},~T^{\pm}.$}

 \par 
 We shall briefly introduce matrix representations 
 for the generators of the Temperley-Lieb algebra. 
 Let us define  operators $e^{\pm}_j$ 
 for $j = 1, \ldots, L-1$ by 
 \begin{equation} 
 e^{\pm}_j = \sum_{a,b,c,d=\pm 1 } {e^\pm}^{ac}_{bd} 
 I_1 \otimes \cdots \otimes I_{j-1}
 \otimes E_j^{ab} \otimes E_{j+1}^{cd} \otimes I_{j+2} 
 \otimes \cdots \otimes I_{L}  \, ,   
 \label{matrix} 
 \end{equation} 
 where the matrix elements ${e^\pm}^{ac}_{bd}$ are given by 
 \begin{equation} 
 {e^\pm}^{ac}_{bd} = 
%
%
    ab e^{\mp(a+b)\lambda/2} \, 
 \delta_{a+c, 0} \, \delta_{b+d,0} 
 \, , \quad {\rm for} \quad a,b,c,d = \pm 1 \, . 
  \label{matrixTL} 
 \end{equation}

 \par 
 Utilizing the matrix representations, 
 we can show that the operators $e^{\pm}_j$'s 
 defined in eq. (\ref{matrix}) commute with the 
 generators $S^{\pm}$ given by (\ref{spm}) in \S 3: 
 \begin{equation} 
 [S^{\pm} , e^+_k]  =  0 \quad {\rm for }   
 \quad   k = 1,2, \ldots, L-1 \, .     
 \label{TL1}
 \end{equation}  
 and
 \begin{equation} 
 [T^{\pm} ,  e^-_k]  =  0 \quad {\rm for }   
 \quad   k = 1,2, \ldots, L-1 \, .     
 \label{TL2}
 \end{equation}  

 In addition we see from (\ref{Xabcd_twist}) 
that $X^{\pm}_j(v)$ can be expressed in
  terms of the Temperley--Lieb operators as
 \begin{equation}
 X_j^{\pm}(v) = 2 \sinh (\lambda-v) \, I + 2\sinh v \,  e^{\pm}_j 
\label{decomp} 
 \end{equation}

 \subsection{Proof of the (anti-)commutation relations 
in the sector $S^z \equiv 0 ~ ({\rm mod}~N)$} 

In the sector $S^z\equiv 0~({\rm mod}~N)$ we need only consider
 untwisted operators with $\phi=0$
 The product of the operators 
 ${X^\pm}_j(v)$'s can be written in terms 
 of the Temperley-Lieb generators ${e^\pm}_j$'s 
 \begin{equation} 
  X^\pm_{L-1}(v)  X^\pm_{L-2}(v) 
 \cdots X^\pm_{1}(v) 
   =   \left(\rho(v)I + f(v) e^\pm_{L-1} \right) 
 \left(\rho(v) I + f(v) e^\pm_{L-2} \right)
 \cdots \left(\rho(v)I + f(v) e^\pm_{1} \right)  
 \end{equation}  
where $\rho(v)=2\sinh(v-\lambda)$ and $f(v)=2\sinh v$ .      
 Thus, using (\ref{TL1})  we  have for all $q$ the commutation relation 
 \begin{equation} 
 [S^{\pm} ,  X^+_{L-1}(v) X^+_{L-2}(v) \cdots  X^+_{1}(v) ] =0 
\label{appcom1} 
\end{equation} 
and 
 \begin{equation}  
 [S^{\pm}, X^+_1(\lambda -v) X^+_2(\lambda -v) 
 \cdots X^+_{L-1}(\lambda -v)] = 0. 
\label{appcom2} 
\end{equation}    
 and for the operators $T^{\pm}$ we use (\ref{TL2}) 
 to obtain 
 \begin{eqnarray} 
 & & [T^{\pm} ,  X^-_{L-1}(v) X^-_{L-2}(v) 
 \cdots X^-_{1}(v) ] =0 \, ,\label{appcom3} \\
 & & [T^{\pm}, X^-_1(\lambda -v) X^-_2(\lambda -v) \cdots
  X^-_{L-1}(\lambda-v)] = 0 \, .  
\label{appcom4}
\end{eqnarray} 
 Then using (\ref{trans}), (\ref{twoterms}) and
 (\ref{appcom1})-(\ref{appcom4})
we find for $S^z\equiv 0 ({\rm mod}~N)$
 the (anti)commutation relations 
 with the transfer matrix  
 \begin{eqnarray} 
 S^{\pm (N)} T(v) & = & q^{N} \, T(v) S^{\pm (N)}   \\
 T^{\pm (N)} T(v)  & = & q^{N} \, T(v) T^{\pm (N)}  \, , 
 \end{eqnarray}
 or equivalently 
 \begin{eqnarray} 
 & & [ S^{\pm (N)},  \,   \Pi_L  T(v)  ] 
 \,  =  0 \\
 & & [ T^{\pm (N)}, \,   \Pi_L  T(v)  ]  \,   =  0 \, .  
 \end{eqnarray}

\subsection{The commutation relations 
in the sector $S^z \equiv n ({\rm mod} N)$} 
%
%

\par 

\par 
We conclude this appendix by considering the commutation 
relation of $(S^+)^n (T^-)^n$  
with the transfer matrix $T(v).$  
From  
(\ref{twoterms}) and (\ref{appcom1}) we have 
\begin{eqnarray} 
\left( \Pi_R T(v; 0) \right) 
(S^+)^n (T^-)^n  
& = & \left( X_{LL}^{+} + \Pi_R X_{RR}^{+} 
\Pi_R \right) (S^+)^n \cdot (T^-)^n  
\nonumber \\ 
& = & \left( (S^+)^n  X_{LL}^{+} + \Pi_R X_{RR}^{+} \Pi_R (S^+)^n \right) 
\cdot (T^-)^n  
\label{decomp1} 
\end{eqnarray} 
We further study the second term in this expression by using 
(\ref{nSR}) twice to obtain for arbitrary $q$ in the sector $S^z=n>0$ 
\begin{eqnarray} 
\Pi_R X_{RR}^{+} 
\left( \Pi_R (S^+)^n \right) (T^-)^n  
& = & \Pi_R X_{RR}^+ 
\left\{ 
(S^{+})^n + q^{n-1} [n] 
(S^{+})^{n-1}S_L^{+} (q^{-2S^z} -1) 
\right\} q^{n \sigma_L^z} 
\Pi_R (T^-)^n
\nonumber \\ 
& = & \Pi_R X_{RR}^+ (S^+)^n q^{n \sigma_L^z} \Pi_R (T^-)^n  \nonumber \\ 
& = & \left(\Pi_R (S^+)^n \right) 
X_{RR}^+ \Pi_R \, q^{n \sigma_1^z} (T^-)^n \nonumber \\ 
& = & 
\left\{ 
(S^{+})^n + q^{n-1} [n] 
(S^{+})^{n-1}S_L^{+} (q^{-2S^z} -1) 
\right\} q^{n \sigma_L^z} \Pi_R 
X_{RR}^+ \Pi_R \, q^{n \sigma_1^z} (T^-)^n  \nonumber \\ 
& = & (S^+)^n \cdot \Pi_R \, q^{n\sigma_1^z} X_{RR}^+ \Pi_R \, 
q^{n\sigma_1^z} (T^-)^n
\label{XRR} 
\end{eqnarray} 
which when used in (\ref{decomp1}) yields 
\begin{eqnarray} 
\left( \Pi_R T(v; 0) \right) (S^+)^n (T^-)^n | n \rangle  
& = & \left( (S^+)^n  X_{LL}^{+} + \Pi_R X_{RR}^{+} \Pi_R (S^+)^n \right) 
\cdot (T^-)^n    
\nonumber \\ 
& = & (S^+)^n   \left( 
X_{LL}^{+} + \Pi_R \, q^{n\sigma_1^z}  X_{RR}^{+} \Pi_R \, q^{n\sigma_1^z} 
\right) 
\cdot (T^-)^n    
\nonumber \\ 
& = & (S^+)^n  \Pi_R T(v; n) (T^-)^n   
\end{eqnarray} 
Similarly we find 
\begin{equation} 
\Pi_R T(v;n)  \cdot  (T^-)^n  
= (T^-)^n  \cdot  \Pi_R T(v;0) . 
\end{equation} 
Thus, we obtain the commutation relation valid for $S^z=n>0$ and all $q$ 
\begin{equation} 
\left[ T(v), (S^+)^n (T^-)^n  
\right] =0 . 
\label{CR_T} 
\end{equation} 
When $q^{2N}=1$ this argument leading to (\ref{CR_T}) immediately extends to 
$S^z\equiv n ({\rm mod}~N).$



\end{document}